\RequirePackage{lineno}
\documentclass[aps,prd,twocolumn,showpacs,superscriptaddress]{revtex4}

\usepackage{graphicx}
\usepackage{dcolumn}
\usepackage{bm}
\usepackage{amsmath}
\usepackage{amssymb}
\usepackage{setspace}

\newcommand{\ppbar}{\ensuremath{p\bar{p}\,}}

\def \pp  {$p\bar{p}$~}
\def \tt  {$t\bar{t}$~}

\def \uu  {$u\bar{u}$~}

\def \bb  {$b\bar{b}$~}

\newcommand{\tbar}{\ensuremath \overline{t}}
\newcommand{\ttbar}{\ensuremath t\tbar}

\newcommand{\ra}{\ensuremath\rightarrow}
\newcommand{\lra}{\ensuremath\leftrightarrow}

\begin{document}

\preprint{PRD draft of FAST : v1.0}

\title{Top quark mass measurement in the \tt all hadronic channel using a matrix element technique in \ppbar ~collisions at $\sqrt {s} = 1.96$~TeV}

\affiliation{Institute of Physics, Academia Sinica, Taipei, Taiwan 11529, Republic of China} 
\affiliation{Argonne National Laboratory, Argonne, Illinois 60439} 
\affiliation{University of Athens, 157 71 Athens, Greece} 
\affiliation{Institut de Fisica d'Altes Energies, Universitat Autonoma de Barcelona, E-08193, Bellaterra (Barcelona), Spain} 
\affiliation{Baylor University, Waco, Texas  76798} 
\affiliation{Istituto Nazionale di Fisica Nucleare Bologna, $^v$University of Bologna, I-40127 Bologna, Italy} 
\affiliation{Brandeis University, Waltham, Massachusetts 02254} 
\affiliation{University of California, Davis, Davis, California  95616} 
\affiliation{University of California, Los Angeles, Los Angeles, California  90024} 
\affiliation{University of California, San Diego, La Jolla, California  92093} 
\affiliation{University of California, Santa Barbara, Santa Barbara, California 93106} 
\affiliation{Instituto de Fisica de Cantabria, CSIC-University of Cantabria, 39005 Santander, Spain} 
\affiliation{Carnegie Mellon University, Pittsburgh, PA  15213} 
\affiliation{Enrico Fermi Institute, University of Chicago, Chicago, Illinois 60637}
\affiliation{Comenius University, 842 48 Bratislava, Slovakia; Institute of Experimental Physics, 040 01 Kosice, Slovakia} 
\affiliation{Joint Institute for Nuclear Research, RU-141980 Dubna, Russia} 
\affiliation{Duke University, Durham, North Carolina  27708} 
\affiliation{Fermi National Accelerator Laboratory, Batavia, Illinois 60510} 
\affiliation{University of Florida, Gainesville, Florida  32611} 
\affiliation{Laboratori Nazionali di Frascati, Istituto Nazionale di Fisica Nucleare, I-00044 Frascati, Italy} 
\affiliation{University of Geneva, CH-1211 Geneva 4, Switzerland} 
\affiliation{Glasgow University, Glasgow G12 8QQ, United Kingdom} 
\affiliation{Harvard University, Cambridge, Massachusetts 02138} 
\affiliation{Division of High Energy Physics, Department of Physics, University of Helsinki and Helsinki Institute of Physics, FIN-00014, Helsinki, Finland} 
\affiliation{University of Illinois, Urbana, Illinois 61801} 
\affiliation{The Johns Hopkins University, Baltimore, Maryland 21218} 
\affiliation{Institut f\"{u}r Experimentelle Kernphysik, Universit\"{a}t Karlsruhe, 76128 Karlsruhe, Germany} 
\affiliation{Center for High Energy Physics: Kyungpook National University, Daegu 702-701, Korea; Seoul National University, Seoul 151-742, Korea; Sungkyunkwan University, Suwon 440-746, Korea; Korea Institute of Science and Technology Information, Daejeon, 305-806, Korea; Chonnam National University, Gwangju, 500-757, Korea} 
\affiliation{Ernest Orlando Lawrence Berkeley National Laboratory, Berkeley, California 94720} 
\affiliation{University of Liverpool, Liverpool L69 7ZE, United Kingdom} 
\affiliation{University College London, London WC1E 6BT, United Kingdom} 
\affiliation{Centro de Investigaciones Energeticas Medioambientales y Tecnologicas, E-28040 Madrid, Spain} 
\affiliation{Massachusetts Institute of Technology, Cambridge, Massachusetts  02139} 
\affiliation{Institute of Particle Physics: McGill University, Montr\'{e}al, Qu\'{e}bec, Canada H3A~2T8; Simon Fraser University, Burnaby, British Columbia, Canada V5A~1S6; University of Toronto, Toronto, Ontario, Canada M5S~1A7; and TRIUMF, Vancouver, British Columbia, Canada V6T~2A3} 
\affiliation{University of Michigan, Ann Arbor, Michigan 48109} 
\affiliation{Michigan State University, East Lansing, Michigan  48824}
\affiliation{Institution for Theoretical and Experimental Physics, ITEP, Moscow 117259, Russia} 
\affiliation{University of New Mexico, Albuquerque, New Mexico 87131} 
\affiliation{Northwestern University, Evanston, Illinois  60208} 
\affiliation{The Ohio State University, Columbus, Ohio  43210} 
\affiliation{Okayama University, Okayama 700-8530, Japan} 
\affiliation{Osaka City University, Osaka 588, Japan} 
\affiliation{University of Oxford, Oxford OX1 3RH, United Kingdom} 
\affiliation{Istituto Nazionale di Fisica Nucleare, Sezione di Padova-Trento, $^w$University of Padova, I-35131 Padova, Italy} 
\affiliation{LPNHE, Universite Pierre et Marie Curie/IN2P3-CNRS, UMR7585, Paris, F-75252 France} 
\affiliation{University of Pennsylvania, Philadelphia, Pennsylvania 19104}
\affiliation{Istituto Nazionale di Fisica Nucleare Pisa, $^x$University of Pisa, $^y$University of Siena and $^z$Scuola Normale Superiore, I-56127 Pisa, Italy} 
\affiliation{University of Pittsburgh, Pittsburgh, Pennsylvania 15260} 
\affiliation{Purdue University, West Lafayette, Indiana 47907} 
\affiliation{University of Rochester, Rochester, New York 14627} 
\affiliation{The Rockefeller University, New York, New York 10021} 
\affiliation{Istituto Nazionale di Fisica Nucleare, Sezione di Roma 1, $^{aa}$Sapienza Universit\`{a} di Roma, I-00185 Roma, Italy} 

\affiliation{Rutgers University, Piscataway, New Jersey 08855} 
\affiliation{Texas A\&M University, College Station, Texas 77843} 
\affiliation{Istituto Nazionale di Fisica Nucleare Trieste/Udine, $^{bb}$University of Trieste/Udine, Italy} 
\affiliation{University of Tsukuba, Tsukuba, Ibaraki 305, Japan} 
\affiliation{Tufts University, Medford, Massachusetts 02155} 
\affiliation{Waseda University, Tokyo 169, Japan} 
\affiliation{Wayne State University, Detroit, Michigan  48201} 
\affiliation{University of Wisconsin, Madison, Wisconsin 53706} 
\affiliation{Yale University, New Haven, Connecticut 06520} 
\author{T.~Aaltonen}
\affiliation{Division of High Energy Physics, Department of Physics, University of Helsinki and Helsinki Institute of Physics, FIN-00014, Helsinki, Finland}
\author{J.~Adelman}
\affiliation{Enrico Fermi Institute, University of Chicago, Chicago, Illinois 60637}
\author{T.~Akimoto}
\affiliation{University of Tsukuba, Tsukuba, Ibaraki 305, Japan}
\author{B.~\'{A}lvarez~Gonz\'{a}lez}
\affiliation{Instituto de Fisica de Cantabria, CSIC-University of Cantabria, 39005 Santander, Spain}
\author{S.~Amerio$^w$}
\affiliation{Istituto Nazionale di Fisica Nucleare, Sezione di Padova-Trento, $^w$University of Padova, I-35131 Padova, Italy} 

\author{D.~Amidei}
\affiliation{University of Michigan, Ann Arbor, Michigan 48109}
\author{A.~Anastassov}
\affiliation{Northwestern University, Evanston, Illinois  60208}
\author{A.~Annovi}
\affiliation{Laboratori Nazionali di Frascati, Istituto Nazionale di Fisica Nucleare, I-00044 Frascati, Italy}
\author{J.~Antos}
\affiliation{Comenius University, 842 48 Bratislava, Slovakia; Institute of Experimental Physics, 040 01 Kosice, Slovakia}
\author{G.~Apollinari}
\affiliation{Fermi National Accelerator Laboratory, Batavia, Illinois 60510}
% ...

\collaboration{CDF Collaboration\footnote{With visitors from $^a$University of Massachusetts Amherst, Amherst, Massachusetts 01003,
$^b$Universiteit Antwerpen, B-2610 Antwerp, Belgium, 
$^c$University of Bristol, Bristol BS8 1TL, United Kingdom,
$^d$Chinese Academy of Sciences, Beijing 100864, China, 
$^e$Istituto Nazionale di Fisica Nucleare, Sezione di Cagliari, 09042 Monserrato (Cagliari), Italy,
$^f$University of California Irvine, Irvine, CA  92697, 
$^g$University of California Santa Cruz, Santa Cruz, CA  95064, 
$^h$Cornell University, Ithaca, NY  14853, 
$^i$University of Cyprus, Nicosia CY-1678, Cyprus, 
$^j$University College Dublin, Dublin 4, Ireland,
$^k$Royal Society of Edinburgh/Scottish Executive Support Research Fellow,
$^l$University of Edinburgh, Edinburgh EH9 3JZ, United Kingdom, 
$^m$Universidad Iberoamericana, Mexico D.F., Mexico,
$^n$Queen Mary, University of London, London, E1 4NS, England,
$^o$University of Manchester, Manchester M13 9PL, England, 
$^p$Nagasaki Institute of Applied Science, Nagasaki, Japan, 
$^q$University of Notre Dame, Notre Dame, IN 46556,
$^r$University de Oviedo, E-33007 Oviedo, Spain, 
$^s$Texas Tech University, Lubbock, TX  79409, 
$^t$IFIC(CSIC-Universitat de Valencia), 46071 Valencia, Spain,
$^u$University of Virginia, Charlottesville, VA  22904,
$^{cc}$On leave from J.~Stefan Institute, Ljubljana, Slovenia, 
}}
\noaffiliation

\date{\today}


\begin{abstract}
We present a measurement of the top quark mass in the all-hadronic channel (\tt $\to$ \bb$q_{1}\bar{q_{2}}q_{3}\bar{q_{4}}$) using 943~pb$^{-1}$ of \ppbar\ collisions at $\sqrt {s} = 1.96$~TeV collected at the CDF II detector at Fermilab (CDF). We apply the standard model production and decay matrix-element (ME) to $\ttbar$ candidate events. We calculate per-event probability densities according to the ME calculation and construct template models of signal and background. The scale of the jet energy is calibrated using additional templates formed with the invariant mass of pairs of jets. These templates form an overall likelihood function that depends on the top quark mass and on the jet energy scale (JES). We estimate both by maximizing this function. Given 72 observed events, we measure a top quark mass of 171.1 $\pm$ 3.7 (stat.+JES) $\pm$ 2.1 (syst.)~GeV/$c^{2}$. The combined uncertainty on the top quark mass is 4.3~GeV/$c^{2}$. 
\end{abstract}


\maketitle

\section{\label{sec:Intro}Introduction}

The top quark plays an important role in particle physics. 
Being the heaviest observed elementary particle results in large contributions to electroweak radiative corrections and provides a constraint on the mass of the elusive Higgs boson. 
More accurate measurements of the top quark mass are important for precision tests of the standard model. 
In addition, having a Yukawa coupling close to unity may indicate a special role for this quark in electroweak symmetry breaking. 
Increasing the precision on the mass of the top quark is central not only for the standard model, but also for other theoretical scenarios beyond the standard model.

At the Tevatron the top quark is produced most frequently via the strong interaction yielding a top/anti-top pair. 
Once produced, the top quark decays into a $b$ quark and a $W$ boson about 99\% of the time according to the standard model. 
Based on the decay mode of the two $W$ bosons the \tt events can be divided in three channels: the dilepton channel when both $W$ bosons decay to leptons; 
the lepton+jets channel when one $W$ boson decays to leptons and the other one decays to hadrons; 
and the all-hadronic channel when both $W$ bosons decay to hadrons. 

This paper reports a measurement of the top quark mass in the all-hadronic channel using 943 pb$^{-1}$ collected with the upgraded CDF II detector at Fermilab. 
In Section~\ref{sec:Det} we give a brief description of the CDF II detector.

The all-hadronic final state consists of six jets, two of which are due to the hadronization of $b$ quarks. 
The large QCD background and jet-parton combinatorics make measurements more difficult in this channel than in the others. 
However, because there are no unobserved final-state particles, it is possible to fully reconstruct all-hadronic events. 
In order to enhance the \tt content over the background, special selection criteria are imposed on the kinematics and topology of the events. 
In Section~\ref{sec:EvSel} we give more details on this selection.

Previous mass measurements of the top quark in the all-hadronic channel were performed at CDF in both Run I~\cite{ahmass1} and Run II~\cite{ahmass2}. 
For the first time in this channel, we measure the mass of the top quark utilizing a technique that uses the matrix element for \tt production and decay. 
The details of the matrix element calculation and implementation are given in Section~\ref{sec:ME}.

The matrix element is used to calculate a probability for each candidate event to be produced via the standard model \tt production mechanism. 
In principle, the mass of the top quark can be determined by maximizing this probability, and such a technique was successfully applied before at CDF in the lepton+jets channel~\cite{lepjets} and in the dilepton channel~\cite{dilepton}. 
In this analysis we take a new approach in that we calculate the matrix element probability in samples of simulated \tt events to build and to parameterize top mass templates. 
These are distributions that depend on the mass of the top quark, unlike the templates for background events whose modeling is described in Section~\ref{sec:Bckd}.
The measured value of the mass of the top quark corresponds to a \tt template whose mixture with a background template best fits the data.
In Section~\ref{sec:MassFit} we give more details on how these templates are built.

Besides considering a matrix element for a different \tt decay channel, in this analysis the matrix element is computed differently than in the aforementioned analyses in the leptonic channels. 
Also, the features of the matrix element probability are exploited to improve the event selection.

The uncertainty on the jet energy scale has the largest contribution to the total uncertainty in most top quark mass measurements. 
In order to minimize this effect, we perform an {\it in situ} calibration of the jet energy scale using the invariant mass of pairs of light flavor jets. 
For \tt events this variable is correlated with the mass of the $W$ boson, and it is sensitive to variations in jet energy scale.
Using this invariant mass we build the dijet mass templates, and we use them for the calibration of the jet energy scale as shown in Section~\ref{sec:MassFit}. 
This procedure, used previously at CDF for the mass measurement of the top quark in the lepton+jets channel~\cite{ljjes}, is used for the first time in the all-hadronic channel in the analysis described in this paper.

The result of the data fit is the topic of Section~\ref{sec:Res}, while in Section~\ref{sec:Syst} the associated systematic uncertainties are described. 
Finally, Section~\ref{sec:Concl} concludes the paper.

\section{\label{sec:Det} Detector}

The CDF II detector is an azimuthally and forward-backward symmetric apparatus designed to study \pp collisions at the Tevatron. 
It is a general purpose detector which combines precision particle tracking with fast projective calorimetry and fine grained muon detection.  

The CDF coordinate system is right handed, with $z$ axis tangent to the Tevatron ring and pointing in the direction of the proton beam. 
The $x$ and $y$ coordinates of a left-handed $x$,$y$, $z$ Cartesian reference system are defined pointing outward and upward from the Tevatron ring, respectively. 
The azimuthal angle $\phi$ is measured relative to the $x$ axis in the transverse plane. 
The polar angle $\theta$ is measured from the proton direction and is typically expressed as pseudorapidity $\eta=-$ln(tan$\frac{\theta}{2}$). 
We define transverse energy as $E_{T} = E$sin$\theta$ and transverse momentum as $p_{T}=p$sin$\theta$ where $E$ is the energy measured in the calorimeter and $p$ is the magnitude of the momentum measured by the tracking system.

Tracking systems are contained in a superconducting solenoid, 1.5~m in radius and 4.8~m in length, which generates a 1.4 T magnetic field parallel to the beam axis. 
The calorimeter surrounds the solenoid. 
A more complete description of the CDF II detector can be found in Ref.~\cite{cdf2}. 
The main features of the detector systems are summarized below. 

The tracking system consists of a silicon microstrip system and an open-cell wire drift chamber that surrounds the silicon. 
The silicon microstrip system consists of eight layers in a cylindrical geometry that extends from a radius of r = 1.35~cm from the beam line to r = 29~cm. 
The layer closest to the beam pipe is a radiation-hard, single sided detector called Layer 00~\cite{l00}. 
The remaining seven layers are radiation-hard, double sided detectors. 
The first five layers after Layer 00 comprise the SVXII~\cite{svx2} system and the two outer layers comprise the ISL~\cite{isl} system. 
This entire system allows track reconstruction in three dimensions. 
The resolution on the impact parameter for high-energy tracks with respect to the interaction point 
is 40~$\mu$m, including a 30~$\mu$m contribution from the beam-line. 
The resolution 
to determine $z_{0}$ ($z$ position of the track at point of minimum distance to interaction vertex) is 70~$\mu$m. 
The 3.1 m long cylindrical drift chamber (COT)~\cite{cot} covers the radial range from 43 to 132~cm and provides 96 measurement layers, organized into alternating axial and $\pm 2^{o}$ stereo superlayers. 
The COT provides full coverage for $|\eta|$ $\le$1. 
The hit position resolution is approximately 140~$\mu$m and the transverse momentum resolution $\sigma(p_{T})/p_{T}^{2}$ = 0.0015~GeV/$c^{-1}$. 


Segmented electromagnetic and hadronic sampling calorimeters surround the tracking system and measure the energy flow of interacting particles in the pseudorapidity range $|\eta|$ $<$ 3.6. 
The central calorimeters (and the end-wall hadronic calorimeter) cover the pseudorapidity range $|\eta|$ $<$ 1.1(1.3) and are segmented in towers of 15$^{o}$ in azimuth and 0.1 in $\eta$. 
The central electromagnetic calorimeter~\cite{cem} uses lead sheets interspersed with polystyrene scintillator as the active medium and photomultipliers. 
The energy resolution for high-energy electrons and photons is $\approx$ 13.5\%/$\sqrt E_{T}\oplus$2\%, where the first term is the stochastic resolution and the second term is a constant term due to the non-uniform response of the calorimeter. 
The central hadronic calorimeter~\cite{cha} uses steel absorber interspersed with acrylic scintillator as the active medium. 
The energy resolution for single-pions is $\approx$ 75\%/$\sqrt E_{T}\oplus$3\% as determined using the test-beam data. 
The plug calorimeters cover the pseudorapidity region 1.1 $<$ $|\eta|$ $<$ 3.6 and are segmented in towers of 7.5$^{o}$ for $|\eta|$ $<$ 2.1 and 15$^{o}$ for $|\eta|$ $>$ 2.1. 
They are sampling scintillator calorimeters coupled with plastic fibers and photomultipliers. 
The energy resolution of the plug electromagnetic calorimeter~\cite{pem} for high-energy electrons and photons is $\approx$ 16\%/$\sqrt E_{T}\oplus$1\%. 
The energy resolution for single-pions in the plug hadronic calorimeter is $\approx$ 74\%/$\sqrt E_{T}\oplus$4\%. 


The collider luminosity is proportional to the average number of inelastic \pp collisions per bunch crossing which is measured using gas \v{C}herenkov counters~\cite{clc} located in the 3.7 $<$ $|\eta|$ $<$ 4.7 region.

The data selection (trigger) and data acquisition systems are designed to accommodate the high rates and large data volume of Run II. 
Based on preliminary information from tracking, calorimetry, and muon systems, the output of the first level of the trigger (level 1) is used to limit the rate of the accepted events to $\approx$ 18~kHz at the luminosity range 3\ra7 $\times$ 10$^{31}$~cm$^{-2}$s$^{-1}$. 
At the next trigger stage (level 2), with more refined information and additional tracking information from the silicon detector, the rate is reduced further to $\approx$ 500~Hz. 
The final level of the trigger (level 3), with access to the complete event information, uses software algorithms and a farm of computers to reduce the output rate to $\approx$ 100~Hz, which is the rate at which events are written to permanent storage.

\section{\label{sec:EvSel} Data Sample and Event Selection}

The expected signature of a \tt event in the all-hadronic channel (\tt $\to$ \bb$q_{1}\bar{q_{2}}q_{3}\bar{q_{4}}$) is the presence of six jets in the reconstructed final state. 
Jets are identified as clusters of energy in the calorimeter using a fixed-cone algorithm with radius 0.4 in $\eta$-$\phi$ space~\cite{jetalg}. 
The energy of the jets needs to be corrected for various effects back to the energy of the parent parton. 
The CDF jet energy corrections are divided into several levels to accommodate different effects that can distort the measured jet energy: non-uniform response of the calorimeter as a function of $\eta$, different response of the calorimeter to different particles, non-linear response of the calorimeter to the particle energies, uninstrumented regions of the detector, multiple \pp interactions, spectator particles, and energy radiated outside the jet clustering cone. 
In this analysis we correct the energy of the jets taking into account all of the above effects except those due to spectator particles and energy radiated outside the cone. 
These additional corrections are recovered using the transfer functions defined in Section~\ref{sec:ME}.

A detailed explanation of the procedure to derive the various individual levels of correction is described in Ref.~\cite{nimjes}. 
Briefly, the calorimeter tower energies are first calibrated as follows. 
The scale of the electromagnetic calorimeter is set using the peak of the dielectron mass resonance resulting from the decays of the $Z$ boson. 
For the hadronic calorimeter we use the single pion test beam data. 
This calibration is followed by a dijet balancing procedure used to determine and correct for variations in the calorimeter response to jets as a function of $\eta$. 
This relative correction ranges from about -10$\%$ to +15$\%$.
After tuning the simulation to reflect the data, a sample of simulated dijet events generated with {\sc pythia}~\cite{pythia} is used to determine the correction that brings the jet energies to the most probable true in-cone hadronic energy. 
The absolute correction varies between 10$\%$ and 40$\%$. 

A systematic uncertainty on these corrections is derived in each case. 
Some are in the form of uncertainties on the energy measurement themselves, and some are uncertainties on the detector simulation. 
Typical overall uncertainty is in the range of 3$\%$ to 4$\%$ for jets with transverse momentum larger than 40 GeV/$c$. 
More details on the estimation of these uncertainties can be found in~\cite{nimjes}.

The data sample is selected using a dedicated multi-jet trigger defined as follows. 
For triggering purposes the calorimeter granularity is simplified to a 24 $\times$ 24 grid in $\eta$-$\phi$ space. 
A trigger tower spans approximately 15$^{o}$ in $\phi$ and 0.2 in $\eta$ covering one or two physical towers. 
At level 1, we require at least one trigger tower with transverse energy $E_{T}^{tow}$ $\geq$ 10~GeV. 
At level 2, we require the sum of the transverse energies of all the trigger towers, $\sum E_{T}^{tow}$, be $\geq$ 175 GeV and the presence of at least four clusters of trigger towers with $E_{T}^{cls}$ $\geq$ 15~GeV.
Finally, at level 3 we require four or more reconstructed jets with $E_{T}$ $\geq$~10 GeV. 
This trigger selects about 80$\%$ of the \tt events in the all-hadronic channel.
The main background present in this data sample is due to the production of multi-jets via QCD couplings.

This analysis relies on Monte Carlo event generation and detector simulation to model the \tt events. We use {\sc herwig} v6.505~\cite{herwig} for the event generation. The CDF II detector simulation~\cite{cdfmc} reproduces the response of the detector to particles produced in \pp collisions. 
Tracking of particles through matter is performed with {\sc geant3}~\cite{geant}. 
Charge deposition in the silicon detectors is calculated using a parametric model tuned to the existing data. 
The drift model for the COT uses the {\sc garfield} package~\cite{garfield}, with the default parameters tuned to match COT data. 
The calorimeter simulation uses the {\sc gflash}~\cite{gflash} parameterization package interfaced with {\sc geant3}. 
The {\sc gflash} parameters are tuned to test-beam data for electrons and pions.
We describe the modeling of the background in Section~\ref{sec:Bckd}. 

The events passing the trigger selection are further required to pass a set of clean-up cuts. 
First, we require the reconstructed primary vertex~\cite{btag} in the event to lie inside the luminous region ($|z|$ $<$ 60~cm). 
In order to reduce the contamination of the sample with events from the leptonic \tt decays, we veto events which have a well identified high-$p_{T}$ electron or muon~\cite{tightlep}, and require that $\frac{\not{E_{T}}}{\sqrt{\sum E_{T}}}$ be $<$ 3~GeV$^{1/2}$~\cite{metsig}, where the missing transverse energy, $\not\!\!{E_{T}}$~\cite{met}, is corrected for both the momentum of any reconstructed muon and the position of the \pp interaction point. The quantity $\sum E_{T}$ is the sum of the transverse energies of jets.

After this preselection, we consider events with exactly six jets, each with transverse energy $E_{T} \geq$ 15~GeV and $|\eta| < $ 2. 
With these six jets, we calculate four variables that are used for the kinematic discrimination of \tt from background.
One of these variables is $\sum E_{T}$ defined above.
Another variable, $\sum_{3}E_{T}$, is the sum of the transverse energies of jets removing the two leading jets.
We define centrality, $C$, as $\frac{\sum E_{T}}{\sqrt{(\sum E)^{2}-(\sum p_{z})^{2}}}$, where $\sum E$ and $\sum p_{z}$ are the sum of the energies of jets and the sum of the momenta of jets along the $z$-axis, respectively. 
The fourth variable is the aplanarity, $A$, defined as $\frac{3}{2}Q_{1}$. 
Here $Q_{1}$ is the smallest normalized eigenvalue of the sphericity tensor $S_{ab}=\sum_{j}P_{a}^{j}P_{b}^{j}$, where $P_{a}^{j}$ is the momentum of a jet along one of the Cartesian axes.
We select events which satisfy the following kinematical cuts: $A$ + 0.005$\sum_{3}E_{T}$ $>$ 0.96, $C$ $>$ 0.78, and $\sum E_{T}$ $>$ 280~GeV.
More details on the clean-up cuts, kinematical and topological variables as well as the optimization of the cuts are given in Ref.~\cite{ahxs}. 

Since the final state of a \tt event is expected to contain two jets originating from $b$ quarks, their identification is important for enhancing the \tt content of our final data sample. 
Jets are identified as $b$ jets using a displaced vertex tagging algorithm. 
This algorithm looks inside the jet for good-quality tracks with hits in both the COT and the silicon detector.
When a displaced vertex can be reconstructed from at least two of those tracks, the signed distance ($L_{2D}$) between this vertex and the primary vertex along the jet direction in the plane transverse to the beams is calculated. 
The jet is considered tagged if $L_{2D}$/$\sigma$($L_{2D}$) $>$ 7.5, where $\sigma$($L_{2D}$) is the uncertainty on $L_{2D}$.
This algorithm has an efficiency of about 60$\%$ for tagging at least one $b$ jet in a simulated \tt event.
More information concerning $b$ tagging is available in Ref.~\cite{btag}. 
In order to improve the signal purity, we require the existence of at least one secondary vertex tag in the event.

\begin{figure}[!htbp]
\begin{center}
\includegraphics[width=\linewidth]{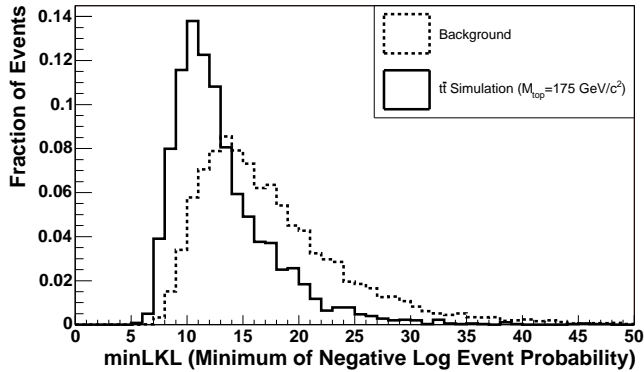}
\caption[Minimum of the negative log event probability]{Distribution of $minLKL$ (minimum of the negative log event probability) for simulated \tt events with $M_{top}$ = 175 GeV/c$^{2}$ (continuous) and for background events (dashed) modeled in Section~\ref{sec:Bckd}. The events pass the kinematical and $b$-tagging requirements.}\label{fig:5-1} 
\end{center}
\end{figure}

We introduce a new variable, $minLKL$, defined as the minimum of the event probability calculated using the matrix element technique (see Section~\ref{sec:ME} for details).
Figure~\ref{fig:5-1} shows the distribution of $minLKL$ for a simulated \tt sample with $M_{top}$ = 175~GeV/$c^{2}$ (continuous line) and for the background (dashed line), after kinematical and $b$ tagging requirements. 
Here and throughout this paper we use $M_{top}$ to label the top quark mass used in the event generation. 
The event probability used in Figure~\ref{fig:5-1} is not normalized due to omission of multiplicative constants in its calculation. 
Although technically this variable is not a probability, we will keep using this name. 
To further reduce the background contribution, we require that $minLKL$ $\le$ 10. 
The value of the cut on $minLKL$ is obtained by minimizing the statistical uncertainty on the top quark mass reconstructed using only the matrix element technique. 
The optimization was done for various top mass quark values using a combination of simulated \tt events and background events (described in Section~\ref{sec:Bckd}). 

Table~\ref{table:5-4} shows the observed number of events in the multi-jet data sample corresponding to a integrated luminosity of 943~pb$^{-1}$ that pass the full event selection. 
The table also shows the expected number of \tt events (S) based on a sample of simulated \tt events assuming the theoretical value of 6.7~pb~\cite{Cacciari} for the \tt production cross section. 
The signal-to-background ratio (S/B) is also shown, where the number of background events (B) is taken as the difference between the observed number of events and the \tt expectation (S).
Based on the results reported in Table~\ref{table:5-4}, the $minLKL$ cut improves the signal-to-background ratio by a factor of three for the sample where only one secondary vertex tag is required, and by a factor of two for the sample where at least two tags are required.
\begin{table}[!htbp]
\begin{center}
\caption{Number of observed multi-jet events passing the event selection corresponding to an integrated luminosity of 943~pb$^{-1}$. The table also shows the expected number of \tt events (S) and the corresponding signal-to-background ratio (S/B). The number of \tt events is based on a sample of simulated \tt events assuming the theoretical value of 6.7~pb~\cite{Cacciari} for the \tt production cross section. The number of background events is taken as the difference between the observed number of multi-jet events and the \tt expectation (S).}\label{table:5-4} 
\begin{tabular*}{\linewidth}{c@{\extracolsep{\fill}}c@{\extracolsep{\fill}}c@{\extracolsep{\fill}}c@{\extracolsep{\fill}}c@{\extracolsep{\fill}}c@{\extracolsep{\fill}}c}
\hline
\hline
Selection & \multicolumn{3}{c}{Single Tag} & \multicolumn{3}{c}{Double Tag} \\
 & Observed & S & S/B & Observed & S & S/B \\
\hline
Kinematical & 782 & 71 & 1/10 & 148 & 47 & 1/2 \\
$minLKL$ $\leq$ 10 & 48 & 13 & 1/3 & 24 & 14 & 1/1 \\
\hline
\hline
\end{tabular*}
\end{center}
\end{table}


\section{\label{sec:ME} Matrix Element Tool}

For each event passing our kinematical and topological requirements, we calculate the corresponding elementary cross section assuming \tt production followed by the all-hadronic decay. 
In this calculation, we consider the momentum 4-vectors of all the observed six jets, but we assume them to be massless. 
The fraction of the total \tt cross section contributed by an event can be interpreted as a probability density for the given event to be part of the \tt production. 
As it is shown in Section~\ref{sec:MEProb}, for each event this probability density depends on the top quark mass. 
The mass value that maximizes the event probability is used in the top quark mass reconstruction technique described in Section~\ref{sec:MassFit}.
A likelihood function obtained by combining the probability densities of a set of events can also be used to reconstruct the top quark mass. 
We use this technique in subsection~\ref{sec:METest} only to validate the matrix element calculation used in the probability density determination, and not for the final mass reconstruction.

\subsection{\label{sec:MEProb} Definition of the probability density}

For any event defined by a set of six 4-momenta, the elementary cross section at a given top quark mass $m$ can be computed as if the event were the result of \tt production followed by all hadronic decay:
\begin{eqnarray}
\label{eq:4-1} d\sigma(m,j) = \int \frac{dz_{a} dz_{b} f(z_{a}) f(z_{b})}{4E_{a}E_{b}|v_{a}-v_{b}|} |\mathcal{M}(m,j)|^{2} \nonumber \\
\times (2\pi)^{4}\delta^{(4)}(E_{F}-E_{I}) \prod_{i=1}^{6} \bigg[ \frac{d^{3}\vec{j}_{i}}{(2\pi)^{3}2E_{i}} \bigg]
\end{eqnarray}
Here, $z_{a}$($z_{b}$) is the fraction of the proton(antiproton) momentum carried by the colliding partons; $f(z_{a})$ and $f(z_{b})$ are the parton distribution functions for protons and for antiprotons respectively; $v_{a}$ and $v_{b}$, and $E_{a}$ and $E_{b}$ represent the velocities and, respectively, the energies of the colliding partons; $j$ is a generic notation for all six 4-momenta in the event assuming perfect parton identification and reconstruction; $\mathcal{M}(m,j)$ is the matrix element corresponding to \tt production and decay in the all hadronic channel; $E_{F}$($E_{I}$) is a generic notation for the 4-vector of the final(initial) state. 

If we sum the elementary cross sections of all the events passing our event selection (trigger, kinematical and topological) without the $minLKL$ requirement, we obtain a fraction ($\epsilon(m)$) of the total \tt cross section, $\sigma_{tot}(m)$:
\begin{equation} 
\label{eq:4-2} \sigma(m) = \int d\sigma(m,j) = \sigma_{tot}(m)\epsilon(m)
\end{equation} 
The fraction $\epsilon(m)$ is equivalent to the event selection efficiency for \tt events and is determined using samples of simulated \tt events. 

For each event, we define the probability density $P(j|m)$ as
\begin{equation}
\label{eq:4-3} P(j|m) = \frac{d\sigma(m,j)}{\sigma_{tot}(m)\epsilon(m) \prod_{i=1}^{6} d^{3}\vec{j}_{i}}
\end{equation}
The quantity $P(j|m) \prod_{i=1}^{6} d^{3}\vec{j}_{i}$ is the probability for an event defined by the set of six jets (i.e., six 4-momenta) to be the result of \tt production followed by an all hadronic decay for top quark mass $m$.

The final state partons from \tt decay are observed as jets in our detector. 
Using simulated \tt events we calculate transfer functions, $TF(\vec{j}|\vec{p})$, which represent a probability for a parton with momentum $\vec{p}$ to be observed as a jet with momentum $\vec{j}$. 
The transfer functions are described in Section~\ref{sec:TF}.

In order to enhance the features of the \tt phase space, an additional weight, $P_{T}(\vec{p})$, is introduced. 
As it is shown in Section~\ref{sec:Pt}, this weight is obtained from the distribution of the transverse momentum of the \tt system in simulated \tt events.

We assume that all six jets present in an all-hadronic \tt event are the result of the hadronization of quarks in the final state.
There is an ambiguity in assigning the jets to the quarks, and therefore all the possible combinations are considered and averaged. 
The counting of all possible assignments is detailed in Section~\ref{sec:combi}. 
The full expression of the probability density is given by:
\begin{eqnarray}
P(j|m) =  \sum_{combi} \int \frac{dz_{a}dz_{b}f(z_{a})f(z_{b})}{4E_{a}E_{b}|v_{a}-v_{b}|} \prod_{i=1}^{6} \bigg[ \frac{d^{3}\vec{p}_{i}}{(2\pi)^{3}2E_{i}} \bigg] \nonumber \\
\label{eq:4-6} \times \frac{|\mathcal{M}(m,p)|^{2} (2\pi)^{4}\delta^{(4)}(E_{F}-E_{I}) TF(\vec{j}|\vec{p}) P_{T}(\vec{p})}{\sigma_{tot}(m)\epsilon(m)N_{combi}}
\end{eqnarray}
where the sum is performed over all jet-parton combinations and $N_{combi}$ is the total number of possible jet-parton assignments.

The calculation of the matrix element $\mathcal{M}(m,p)$ is detailed in Section~\ref{sec:mecalc} and the integration performed in Eq.~\ref{eq:4-6} is described in Section~\ref{sec:integr}.

\subsubsection{\label{sec:mecalc}Calculation of the matrix element}
 
Two processes contribute to \tt production: gluon-gluon fusion and quark-antiquark annihilation. 
At the Tevatron, about $(15\pm5)\%$ of \tt events are expected to be produced by gluon-gluon fusion while the remaining $85\%$ are produced by quark-antiquark annihilation~\cite{Cacciari}. 
In addition, $90\%$ of the simulated \tt events produced by quark-antiquark annihilation result from \uu annihilation. 
Given that having both types of \tt production doubles the calculation time, we only use the matrix element describing the process \uu$\to$\tt$\to$ $(bu\bar{d})(\bar{b}\bar{u}d)$. 
To validate this choice, we reconstruct the top quark mass using a \uu matrix element in a sample of \tt events produced only via gluon-gluon fusion. 
We observe a negligible bias (0.0 $\pm$ 0.2 GeV/$c^{2}$) in the reconstruction of the top quark mass and we conclude that using a matrix element with \uu as the initial state should be sufficient for the mass reconstruction.

For the final state, having a $W$ boson decay into a $u\bar{d}$ pair or a $c\bar{s}$ pair results in no difference for the final reconstruction as both pairs of quarks will be observed as jets. 
The other hadronic decays are suppressed since their rate is proportional to the square of small elements of the Cabibbo-Kobayashi-Maskawa matrix~\cite{ckm}. 

In the high-energy limit (or the massless limit), the solutions to the Dirac equation can be written as 
\begin{eqnarray}
 u(p) = \sqrt{2E_{p}}
\left ( \begin{array}{r}
\frac{1}{2} (1-\hat{p} \cdot \vec{\sigma}) \xi \\
\frac{1}{2} (1+\hat{p} \cdot \vec{\sigma}) \xi 
\end{array} \right ) \nonumber \\
\label{eq:4-15} v(p) = \sqrt{2E_{p}}
\left ( \begin{array}{r}
\frac{1}{2} (1-\hat{p} \cdot \vec{\sigma}) \eta \\
-\frac{1}{2} (1+\hat{p} \cdot \vec{\sigma}) \eta 
\end{array} \right )
\end{eqnarray}
where $p=(E_{p},\vec{p})$ is the 4-momentum of a particle. 
The solution with positive frequencies is $u(p)$, and that with negative frequencies is $v(p)$; $\sigma^{\mu}=(1,\vec{\sigma})$ and $\overline{\sigma}^{\mu}=(1,-\vec{\sigma})$, where $\vec{\sigma}$ are the Pauli spin matrices.

The presence of the operator $\hat{p} \cdot \vec{\sigma}$ will project the spin states along the direction of motion defined by $\hat{p}$. For a particle traveling in the direction defined by the polar angle $\theta$ and by the azimuthal angle $\phi$, the spin states along this direction are given by Eq.~\ref{eq:4-16}.
\begin{equation}
\label{eq:4-16} \xi(\uparrow) = 
\left ( \begin{array}{c}
\cos \frac{\theta}{2} \\
e^{i\phi} \sin \frac{\theta}{2}
\end{array} \right )
, \xi(\downarrow) =
\left ( \begin{array}{c}
-e^{-i\phi} \sin \frac{\theta}{2} \\
\cos \frac{\theta}{2}
\end{array} \right )
\end{equation}

For an antiparticle we have $\eta(\uparrow)=\xi(\downarrow)$ and $\eta(\downarrow)=-\xi(\uparrow)$. Given these relations and assuming that the incoming partons travel along the $z$-axis, the matrix element has only two non-zero terms due to the initial state partons, $I_{RR}$ and $I_{LL}$. These are 4-vectors and correspond to the situations when the incoming partons have the same chirality. Considering the proton going in the positive direction, these terms are:
\begin{eqnarray}
I_{RR}^{\mu} & = & \sqrt{2E_{u}^{in}} \sqrt{2E_{\overline{u}}^{in}} \; (0,1,i,0) \nonumber \\
\label{eq:4-20} I_{LL}^{\mu} & = & \sqrt{2E_{u}^{in}} \sqrt{2E_{\overline{u}}^{in}} \; (0,1,-i,0)
\end{eqnarray}
where $E_{u}^{in}$ and $E_{\overline{u}}^{in}$ are the energies of the incident $u$ quark and $\overline{u}$ quark, respectively. 

Omitting all multiplicative constants, we express the matrix element squared as
\begin{eqnarray}
|\mathcal{M}|^{2} \to \sum_{\substack{spins \\ colors}} |\mathcal{M}|^{2} = F_{E}^{2} \cdot \widetilde{P}_{g} \cdot \widetilde{P}_{t} \cdot \widetilde{P}_{\overline{t}} \cdot \widetilde{P}_{W_{1}} \nonumber \\
\label{eq:4-24} \times \widetilde{P}_{W_{2}} \cdot (|M_{RR}|^{2}+|M_{LL}|^{2})
\end{eqnarray}
where the factors entering this expression are 
\begin{equation}
\label{eq:4-25} \left \{ \begin{array}{l} 
F_{E} = \sqrt{(E_{b}) (E_{\overline{b}}) (E_{u}) (E_{\overline{d}}) (E_{d}) (E_{\overline{u}}) (E_{u}^{in}) (E_{\overline{u}}^{in})} \\
\widetilde{P}_{g} = |P_{g}|^2 = \frac{1}{(p_{u}+p_{\overline{u}})^{4}} \\
\widetilde{P}_{t,\overline{t}} = \frac{1}{(p_{t,\overline{t}}^{2} - m^{2})^{2} + m^{2} \Gamma_{t}^{2}} \\
\widetilde{P}_{W_{1,2}} = \frac{1}{(P_{W_{+,-}}^{2} - M_{W}^{2})^{2} + M_{W}^{2} \Gamma_{W}^{2}} \\
M_{I} = \left (
\begin{array}{cc}
\xi_{b}^{\dagger}(\downarrow) \, (W_{-} \cdot \overline{\sigma}) & 0
\end{array} \right )
S_{I}
\left ( \begin{array}{c}
0 \\
(W_{+} \cdot \overline{\sigma}) \, \xi_{\overline{b}}^{\dagger}(\downarrow)
\end{array} \right )
\end{array} \right.
\end{equation}
In Eq.~\ref{eq:4-25} $M_{W}$ and $\Gamma_{W}$ are the mass and the width of the $W$ boson, $m$ and $\Gamma_{t}$ are the mass and the width of the top quark, and $W_{+}$ and $W_{-}$ are the 4-momenta of the $W$ bosons. 

Also in Eq.~\ref{eq:4-25} $M_{I}$ stands for both $M_{RR}$ and $M_{LL}$ (Eq.~\ref{eq:4-24}), the difference arising from the definition of the symbol $S_{I}$. 
The symbol $S_{I}$ is defined as 
\begin{equation}
\label{eq:4-26}S_{I}=p_{t}^{\mu} \gamma_{\mu} I^{\nu} \gamma_{\nu} p_{\overline{t}}^{\rho} \gamma_{\rho} + m^{2} I^{\mu} \gamma_{\mu}
\end{equation}
where $\gamma_{\mu}$ are the Dirac matrices and $I$ is either $I_{RR}$ or $I_{LL}$. 
We calculate $M_{RR}$ and $M_{LL}$ using Eq.~\ref{eq:4-16} and matrix algebra~\cite{peskin}.

\subsubsection{\label{sec:combi}Combinatorics}
While there are $6!=720$ ways to assign the observed jets to the six partons of the final state in all hadronic \tt decay, we can take into account a reduced number of possibilities by making a few observations and assumptions. 

In the case of the process \uu $\to$ \tt, assuming that the masses of the up quarks are negligible and omitting the constant and the gluon propagator terms, the spin averaged matrix element squared is
\begin{equation}
\label{eq:4-9} \frac{1}{4} \sum_{spins} |\mathcal{M}|^{2} \approx (p_{u} \cdot p_{\overline{t}}) (p_{\overline{u}} \cdot p_{t}) +  (p_{u} \cdot p_{t}) (p_{\overline{u}} \cdot p_{\overline{t}}) + m (p_{u} \cdot p_{\overline{u}}) 
\end{equation}
where $p$ is the 4-momentum of a particle.

From this expression, the $t \leftrightarrow \overline{t}$ symmetry is evident. 
The symmetry holds also for the matrix element of the process containing the decay of the top quarks. 
This symmetry can be translated into a symmetry to $b \leftrightarrow \overline{b}$ once we consider all possible $b$-$W$ pairings for each top quark: $\{ t=(b_{1},W_{1}), \overline{t}=(b_{2},W_{2}) \}, \{ t=(b_{1},W_{2}), \overline{t}=(b_{2},W_{1}) \}$. It is obvious that swapping the $b$'s is equivalent with swapping the top quarks.

In conclusion, due to the $t \leftrightarrow \overline{t}$ symmetry the number of relevant combinations is $360$. 
Secondly, if any of the jets is identified as a secondary vertex tag we assume that jet be produced by a $b$ quark. 
This assumption results in a factor of three reduction of the number of relevant combinations, down to $120$ (or $5!$). 
If there is an additional secondary vertex tag, we get a factor of five reduction down to $24$ (or $4!$). 
If there are more than two secondary vertex tags, we assign to $b$ quarks only the two jets with the highest transverse energy.
Note that the quarks in the decay of either $W$ boson can not be interchanged in the matrix element calculation as one is particle and the other is antiparticle and they have different spinors.

\subsubsection{\label{sec:TF}Transfer functions}

The transfer functions, $TF(\vec{j}|\vec{p})$, express the probability for a parton with momentum $p$ to be associated with a jet reconstructed to have momentum $j$.
The transfer function term from Eq.~\ref{eq:4-6} is in fact a product of six terms, one for each of the final state quarks: two for the $b$ quarks and four for the decay products of the $W$ bosons. 
For each jet in the final state we assume that the jet axis is the same as that of the parton that went on to form the jet. 
Making the  change of variables $j \to \zeta=1-j/p$, the expression for $TF(\vec{j}|\vec{p})$ becomes:
\begin{eqnarray}
\label{eq:4-30} TF(\vec{j}|\vec{p}) \to \prod_{i=1}^{6} TF(\vec{j_{i}}|\vec{p_{i}}) = \prod_{i=1}^{6} \widetilde{TF}(\zeta(j_{i})|p_{i}) \nonumber \\
\times \frac{(-1)}{p_{i}} \; \delta^{(2)}(\Omega_{J_{i}}-\Omega_{P_{i}})
\end{eqnarray}
where $\Omega_{J_{i}}$ and $\Omega_{P_{i}}$ are the solid angles of the jets and of the quarks, respectively. 
The transfer functions $\widetilde{TF}(\zeta_{i}|p_{i})$ are built using simulated \tt events with $M_{top}$ = 175~GeV/$c^{2}$ surviving the trigger, kinematical and topological requirements. 
The choice of $M_{top}$ = 175~GeV/$c^{2}$ is arbitrary as our studies show that the transfer functions have a negligible dependence on the mass of the top quark in the range 150 GeV/$c^{2}$ to 200 GeV/$c^{2}$. 
In this sample, we associate a jet with a parton if their separation in the $\eta-\phi$ space is $\Delta R=\sqrt{\Delta\eta^{2} + \Delta\phi^{2}} \leq 0.4$. 
We define a jet to be matched to a parton if no other jet satisfies this geometrical requirement. 
We define a \tt event to be matched if each of the six partons in the final state has a unique jet matched to it. 
The transfer functions are built out of the sample of matched events. 

The jets formed by partons from $W$-bosons decays have a different energy spectrum from that of the jets originating from the $b$ quarks. 
Thus we form different sets of transfer functions depending on the flavor of the parton the jet has been matched to.

The transfer functions are described using a parameterization in bins of the parton energies and of the parton pseudorapidities. 
We use three bins for the pseudorapidity: $0 \lra 0.7$, $0.7 \lra 1.3$, and $1.3 \lra 2.0$.
Table~\ref{table:4-2} shows the definition of energy binning for the $b$-jet transfer functions, while Table~\ref{table:4-3} is for the $W$-jet transfer functions. 
The energy binning is chosen such that the distributions for transfer functions are smooth.
In each bin, the shape of the transfer function is fitted to a normalized sum of two Gaussians.

\begin{table}[!htbp]
\begin{center}
\caption[Definition of the binning of the parton energy for $b$-jet]{Definition of the binning of the parton energy for the $b$-jet transfer functions parameterization for various $\eta$ bins. The unit for the energy values is~GeV.}\label{table:4-2}
\begin{tabular*}{\linewidth}{c@{\extracolsep{\fill}}c@{\extracolsep{\fill}}c@{\extracolsep{\fill}}c}
\hline
\hline
Bin & $0 \le |\eta| < 0.7$ & $0.7 \le |\eta| < 1.3$ & $1.3 \le |\eta| \le 2.0$ \\
\hline
1 & $10 \to 53$ & $10 \to 83$ & $10 \to \infty$ \\
2 & $53 \to 64$ & $83 \to 111$ & \\
3 & $64 \to 74$ & $111 \to \infty$ & \\
4 & $74 \to 85$ & & \\
5 & $85 \to 97$ & & \\
6 & $97 \to 114$ & & \\
7 & $114 \to \infty$ & & \\
\hline
\hline
\end{tabular*}
\end{center}
\end{table}

\begin{table}[!htbp]
\begin{center}
\caption[Definition of the binning of the parton energy for $W$-jet]{Definition of the binning of the parton energy for the $W$-jet transfer functions parameterization for various $\eta$ bins. The unit for the energy values is~GeV.}\label{table:4-3}
\begin{tabular*}{\linewidth}{c@{\extracolsep{\fill}}c@{\extracolsep{\fill}}c@{\extracolsep{\fill}}c}
\hline
\hline
Bin & $0 \le |\eta| < 0.7$ & $0.7 \le |\eta| < 1.3$ & $1.3 \le |\eta| \le 2.0$\\
\hline
1 & $10 \to 32$ & $10 \to 50$ & $10 \to 98$ \\
2 & $32 \to 38$ & $50 \to 63$ & $98 \to \infty$\\
3 & $38 \to 44$ & $63 \to 76$ & \\
4 & $44 \to 49$ & $76 \to 90$ & \\
5 & $49 \to 54$ & $90 \to 108$ & \\
6 & $54 \to 59$ & $108 \to \infty$ & \\
7 & $59 \to 64$ & & \\
8 & $64 \to 69$ & & \\
9 & $69 \to 75$ & & \\
10 & $75 \to 81$ & & \\
11 & $81 \to 89$ & & \\
12 & $89 \to 99$ & & \\
13 & $99 \to 113$ & & \\
14 & $113 \to \infty$ & & \\
\hline
\hline
\end{tabular*}
\end{center}
\end{table}

\subsubsection{\label{sec:Pt}Transverse momentum of the \tt system}

The $P_{T}(\vec{p})$ weight (introduced in Eq.~\ref{eq:4-6}) is a function dependent on the momenta of the partons in the final state, generically represented by $\vec{p}$ in the argument of the function. 
More exactly, this weight depends on the magnitude of the transverse momentum of the \tt system, $p_{T}^{t\bar{t}}$, and azimuthal angle, $\phi_{T}^{t\bar{t}}$. 
As we expect to have a flat dependence on $\phi_{T}^{t\bar{t}}$ we express this through a factor of $1/2\pi$. 
We define the function $\widetilde{P}_{T}(p_{T}^{t\bar{t}})$ to express the dependence on $p_{T}^{t\bar{t}}$.
We write in Eq.~\ref{eq:4-35} the expression of the weight due to the transverse momentum of the \tt system.
\begin{equation}
\label{eq:4-35} P_{T}(\vec{p}) \to P_{T}(p_{x}^{t\bar{t}},p_{y}^{t\bar{t}}) = \frac{\widetilde{P}_{T}\Big(p_{T}^{t\bar{t}} = \sqrt{(p_{x}^{t\bar{t}})^{2}+(p_{y}^{t\bar{t}})^{2}}\Big)}{2\pi\sqrt{(p_{x}^{t\bar{t}})^{2}+(p_{y}^{t\bar{t}})^{2}}} 
\end{equation}
where $p_{T}^{t\bar{t}}$ is shown in its Cartesian form using the projections of the transverse momentum of the \tt system along the $x$ and $y$ axes. 

\begin{figure}[!htbp]
\begin{center}
\includegraphics[width=\linewidth]{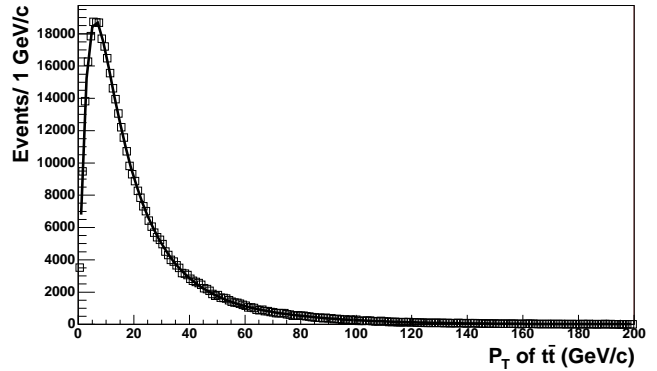}
\end{center}
\caption[Transverse momentum of the $t\overline{t}$ events]{Magnitude of the transverse momentum of the \tt system in simulated \tt events. The fit is a sum of three Gaussians.}\label{fig:4-4}
\end{figure}

Using the same simulated \tt events as for transfer functions, in Fig.~\ref{fig:4-4} we show the distribution of the magnitude $p_{T}^{t\bar{t}}$ of the transverse momentum of the \tt system. A sum of three Gaussians is a good fit of this distribution. 

\subsubsection{\label{sec:integr}Implementation and evaluation of the probability density}

The Sections~\ref{sec:combi}, ~\ref{sec:mecalc}, ~\ref{sec:TF} and ~\ref{sec:Pt} present details on the expressions of several important pieces entering the probability density. 
To carry out the integration over parton momenta, we change to a spherical coordinate system. 
The delta functions $\delta^{(2)}(\Omega_{J_{i}}-\Omega_{P_{i}})$ present in the expression of the transfer functions $TF(\vec{j}|\vec{p})$(Eq.~\ref{eq:4-30}) allow us to drop all integrals over the parton angles.

To further reduce the number of integrals we use the narrow width approximation for the $W$ bosons. 
This results in two more delta functions for the squares of the propagators of the two $W$ bosons exemplified in Eq.~\ref{eq:4-38} for both bosons. 
\begin{eqnarray}
\label{eq:4-38} \widetilde{P}_{W} = \frac{1}{(P_{W}^{2} - M_{W}^{2})^{2} + M_{W}^{2} \Gamma_{W}^{2}} \stackrel{\Gamma_{W} \ll M_{W}}{\longrightarrow} \nonumber \\
\longrightarrow \delta(P_{W}^{2} - M_{W}^{2}) \frac{\pi}{M_{W} \Gamma_{W}}
\end{eqnarray}

In the high-energy limit, the invariant mass of the $W^{+}$-boson decay products is given by: 
\begin{eqnarray}
\label{eq:4-39} P_{W^{+}}^{2} = 2 p_{1} p_{2} \sin \theta_{1} \sin \theta_{2} (\cosh \Delta \eta_{12} - \cos \Delta \phi_{12}) = \nonumber \\
= 2 p_{1} p_{2} \omega_{12} (\Omega_{1},\Omega_{2})
\end{eqnarray}
where $\Delta \eta_{12}$ is the difference in pseudorapidities of the two decay partons and $\Delta \phi_{12}= \pi - ||\phi_{1}-\phi_{2}| - \pi|$ is the difference between their azimuthal angles. 

Making the change of variables $P_{W^{+}}^{2} \to p_{1}$, Eq.~\ref{eq:4-38} can be written as: 
\begin{equation}
\label{eq:4-40} \widetilde{P}_{W^{+}} \stackrel{\Gamma_{W} \ll M_{W}}{\longrightarrow} \frac{\pi}{M_{W} \Gamma_{W}} \frac{1}{2 p_{2} \omega_{12} (\Omega_{1},\Omega_{2})} \delta(p_{1} - p_{1}^{0}) 
\end{equation}
where $p_{1}^{0}=M_{W}^{2}/(2 p_{2} \omega_{12})$. 
In the case of the $W^{-}$ boson we use equations similar to Eqs.~\ref{eq:4-39} and~\ref{eq:4-40}, but with different notations: the change of variables is $P_{W^{-}}^{2} \to p_{3}$ and the pole of the delta function is $p_{3}^{0}=M_{W}^{2}/(2 p_{4} \omega_{34})$. 
The mass and width of the $W$ boson are fixed at 80.4~GeV/$c^{2}$ and 2.1~GeV/$c^{2}$, respectively~\cite{pdg}.

As described in section~\ref{sec:mecalc}, we assume that the incoming partons have zero transverse momentum. 
This would, in principle, result in violation of momentum conservation in the transverse plane as we consider non-zero transverse momentum for the \tt system in the ME calculation. However, we expect this to be a small effect covered by the uncertainty on the parton distribution functions of the proton and of the antiproton. We can omit the delta functions requiring energy conservation along the $x$ and $y$ axes, resulting in
\begin{eqnarray}
\delta^{(4)}(E_{F}-E_{I}) \to \delta \big(E_{a}+E_{b} - \sum_{i=1}^{6}p_{i} \big) \nonumber \\
\times \delta \big(p_{a}^{z}+p_{b}^{z} - \sum_{i=1}^{6}p_{i}^{z} \big) = \nonumber \\
\label{eq:4-42} = \delta \big(p_{u}+p_{\overline{u}} - \sum_{i=1}^{6}p_{i} \big) \; \delta \big(p_{u}-p_{\overline{u}} - \sum_{i=1}^{6}p_{i}^{z}\big)
\end{eqnarray}

We make the change of variables $z_{a} \to p_{u}$ and $z_{b} \to p_{\overline{u}}$ since $z_{a}=p_{u}/p_{proton}$ and $z_{b}=p_{\overline{u}}/p_{antiproton}$. 
The values of the proton and antiproton momenta, $p_{proton}$ and $p_{antiproton}$, are constant and from now on we drop them from any expressions. 
In the high-energy limit we have $|v_{a}-v_{b}|=2c$ and we omit this term since $c$ is a constant, the speed of light. 
We express the energy-conserving delta function as
\begin{eqnarray}
\delta^{(4)}(E_{F}-E_{I}) \to \delta \big(p_{u}+p_{\overline{u}} - \sum_{i=1}^{6}p_{i} \big) \nonumber \\
\times \delta \big(p_{u}-p_{\overline{u}} - \sum_{i=1}^{6}p_{i}\cos \theta_{i}\big) = \nonumber \\
\label{eq:4-43} = \frac{1}{2} \delta (p_{u}-p_{u}^{0}) \delta (p_{\overline{u}}-p_{\overline{u}}^{0} )
\end{eqnarray}
where $p_{u}^{0} = \sum_{i=1}^{6}p_{i}(1 + \cos \theta_{i} )/2$ and $p_{\overline{u}}^{0} = \sum_{i=1}^{6}p_{i}(1 - \cos \theta_{i})/2$.
 
In section~\ref{sec:Pt}, we expressed $P_{T}(\vec{p})$ as a function of the projections of the transverse momentum of the \tt system along the $x$ and $y$ axes (Eq.~\ref{eq:4-35}). We will make a change of variable from the $b$-quark momenta to these variables. The Jacobian of this transformation
\begin{equation}
\label{eq:4-45} J(b \to 6) = \frac{1}{\sin \theta_{b} \sin \theta_{\overline{b}} (\cos \phi_{b}  \sin \phi_{\overline{b}}  - \sin \phi_{b}  \cos \phi_{\overline{b}})}
\end{equation}
is obtained by solving the system of equations for $p_{b}$ and $p_{\overline{b}}$. 
\begin{equation}
\label{eq:4-46} \left \{ \begin{array}{l}
p_{x}^{t\bar{t}} = p_{b} \cos \phi_{b} \sin \theta_{b} + p_{\overline{b}} \cos \phi_{\overline{b}} \sin \theta_{\overline{b}} + \sum_{i=3}^{6}p_{i}^{x} \\
p_{y}^{t\bar{t}} = p_{b} \sin \phi_{b} \sin \theta_{b} + p_{\overline{b}} \sin \phi_{\overline{b}} \sin \theta_{\overline{b}} + \sum_{i=3}^{6}p_{i}^{y} 
\end{array} \right.
\end{equation}

We write the expression of the probability density in its final form as
\begin{eqnarray}
P(j|m) =  \sum_{combi}  \int \frac{dp_{x}^{t\bar{t}}dp_{y}^{t\bar{t}}dp_{2}dp_{4}}{\sigma_{tot}(m)\epsilon(m)N_{combi}} \nonumber \\
\label{eq:4-47} \times \frac{J(b \to 6) p_{b} p_{\overline{b}} f(p_{u}^{0}) f(p_{\overline{u}}^{0})}{(\omega_{12})^{2}(\omega_{34})^{2}p_{2}p_{4}} \prod_{i=1}^{6} \bigg[ \widetilde{TF}(\zeta_{i}|p_{i}) \bigg] \nonumber \\
\times \frac{\widetilde{P}_{T}(p_{T}^{t\bar{t}})}{p_{T}^{t\bar{t}}} \cdot \widetilde{P}_{g} \cdot \widetilde{P}_{t} \cdot \widetilde{P}_{\overline{t}} \cdot (|M_{RR}|^{2}+|M_{LL}|^{2}) 
\end{eqnarray}
We evaluate the integrals in Eq.~\ref{eq:4-47} numerically.
The integration is performed in the interval $[-60,60]$~GeV/$c$ for the variables $p_{x,y}^{t\bar{t}}$ and $[10,300]$~GeV/$c$ for the variables $p_{2,4}$. 
The step of integration is 2~GeV/$c$. 
Based on a sample of \tt events where $M_{top}$ = 175~GeV/$c^{2}$ passing the event selection, we choose these integration ranges such that the distributions of the parton level variables ($p_{x,y}^{t\bar{t}}$ and momenta of $W$-boson decay partons) are contained well (99$\%$) within them. 
Given these limits, at each step of integration we have to make sure that all momenta entering Eq.~\ref{eq:4-47} have positive magnitudes. 
The probability density is evaluated for top mass values going in 1~GeV/$c^{2}$ increments from 125~GeV/$c^{2}$ to 225~GeV/$c^{2}$.

\begin{figure}[!htbp]
\begin{center}
\includegraphics[width=\linewidth]{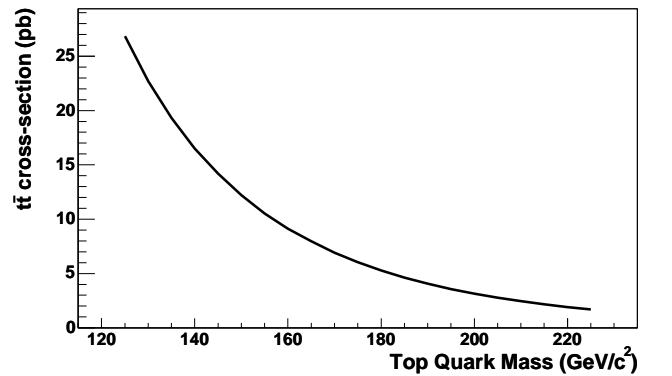}
\end{center}
\caption[Cross section for $t\overline{t}$ production versus the top mass, from CompHep]{Cross section for $t\overline{t}$ production as a function of the top quark mass, as obtained from {\sc comphep}~\cite{comphep}.}\label{fig:4-3}
\end{figure}

The dependence on mass of the $t\overline{t}$ cross section is obtained from values calculated at leading-order by {\sc comphep}~\cite{comphep} Monte Carlo generator for the processes $u\overline{u} \to t\overline{t}$, $d\overline{d} \to t\overline{t}$ and $gg \to t\overline{t}$. 
The absolute values for these cross sections are not as important as their top mass dependence, which is shown in Fig.~\ref{fig:4-3}.

For the proton and antiproton parton distribution functions (PDF), $f(p_{u}^{0}) f(p_{\overline{u}}^{0})$, we use the {\sc cteq5l}~\cite{cteq5l} distributions with the scale corresponding to a top mass of 175~GeV/$c^{2}$. The $t\overline{t}$ acceptance, $\epsilon(m)$, is described in Section~\ref{sec:MEProb}.


\subsection{\label{sec:METest} Validation of the  matrix element calculation}

The event probability described in the Section~\ref{sec:MEProb} is expected to have a maximum around the true top quark mass in the event. 
Multiplying all the event probabilities we obtain a likelihood function,
\begin{equation}
\label{eq:4-48} L(M_{top}) = \prod_{events} P(j|M_{top})
\end{equation}
which is expected to have a maximum around the true top quark mass of the sample.
Finding the value of the top quark mass that maximizes the likelihood represents the traditional method for reconstructing the top quark mass using a matrix element technique~\cite{dilepton}. 
However, we use this reconstruction technique only to check the matrix element calculation.

We use the simulated \tt samples generated with various top quark masses. 
For each sample, we reconstruct the top quark mass using the traditional matrix element technique and compare the reconstructed mass to the true input mass $M_{top}$ for several different input mass values.
Ideally, we should see a linear dependence with no bias and a unit slope.

\begin{figure}[!htbp]
\begin{center}
\includegraphics[width=\linewidth]{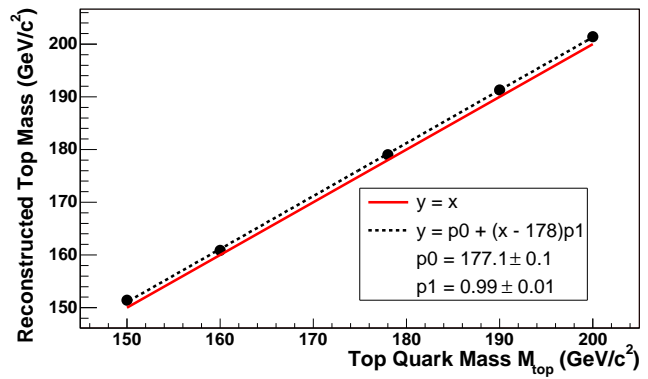}
\end{center}
\caption[Mass reconstruction using smeared parton energies]{Reconstructed top mass versus input top mass at parton level. The energies of the partons have been smeared using the transfer functions. The continuous line $y=x$ is added for visual reference.}\label{fig:4-5}
\end{figure}

The first check is done at the parton level. 
We smear the energies of the final state partons from our simulated \tt events and use these numbers to describe the jets. 
The parton energies are smeared according to the transfer functions described in section~\ref{sec:TF}.
Figure~\ref{fig:4-5} shows the linearity check in this case. 
We observe a slope of $\approx$1 and a bias of 0.9~GeV/$c^{2}$.

\begin{figure}[!htbp]
\begin{center}
\includegraphics[width=\linewidth]{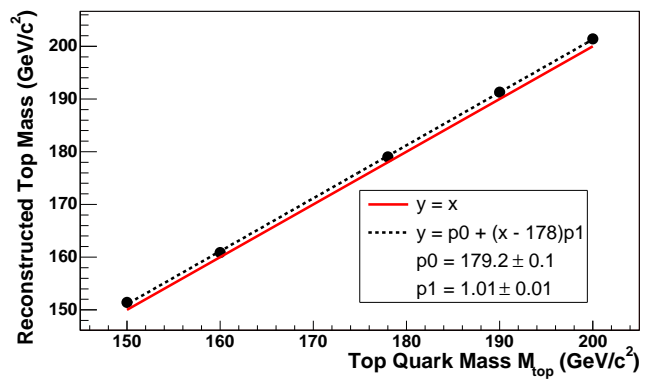}
\end{center}
\caption[Mass reconstruction using jets matched to partons]{Reconstructed top mass versus input top mass using jets that were uniquely matched to partons. The continuous line $y=x$ is added for visual reference.}\label{fig:4-6}
\end{figure}

\begin{figure}[!htbp]
\begin{center}
\includegraphics[width=\linewidth]{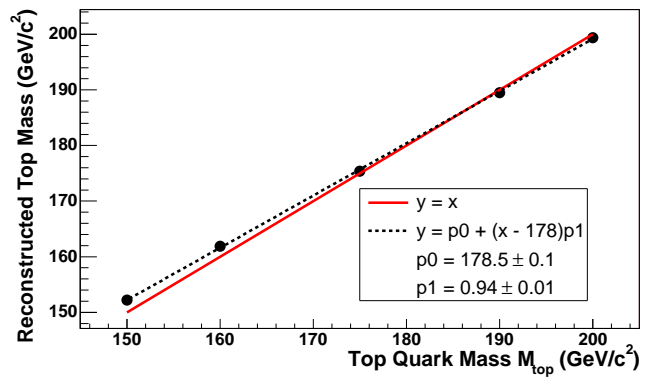}
\end{center}
\caption{Reconstructed top mass versus input top mass using realistic jets. The continuous line $y=x$ is added for visual reference.}\label{fig:4-7}
\end{figure}

We perform the same test using the energies of the jets matched to the partons. 
Figure~\ref{fig:4-6} shows the linearity check. 
Here the bias is 1.2~GeV/$c^{2}$, but the slope remains $\approx$1. 
The final test we perform to validate the matrix element calculation uses fully reconstructed signal events where we allow events to include mismatched jets as well.
Figure~\ref{fig:4-7} shows the linearity check in this case. 
The bias is no longer the same for all masses as the slope is 0.94 $\pm$ 0.01. 

Although there is some bias, all checks we list above show the good performance of our matrix element calculation. 
In general, the traditional matrix element approach~\cite{lepjets} is expected to provide a better statistical uncertainty on the top mass than the template analyses~\cite{ljjes}. 
In the case of the present analysis, our studies show that the traditional matrix element method does better only when the mass reconstruction is performed on signal samples. 
When the background is mixed in, the template method we use has a greater sensitivity and by construction eliminates the bias of the matrix element calculation (see Section~\ref{sec:MassFit}).



\section{\label{sec:Bckd} Background Model}

In this section we describe the data-driven technique used to model the background for this analysis. 
The technique uses jet energies which are measured in the calorimeter and so are unchanged by jet energy scale changes. 
Properties of the model are checked by comparison with a simulated sample of events containing the final state $b\overline{b}$ + 4 light partons. 

The modeling of background is based on a subset of the multi-jet data sample depleted of \tt events where the heavy flavor jets are identified according to background-like heavy flavor rates (tagging matrix), described in Section~\ref{sec:mistag}. 
The subset of multi-jet data is selected applying the event selection of Section~\ref{sec:EvSel} excluding the $minLKL$ and the secondary vertex tag requirements. 
This sample (BG) counts 2652 events, with an estimated signal-to-background ratio of about 1/25.
For this ratio we estimate the signal from a sample of simulated \tt events assuming a \tt production cross section of 6.7~pb. 
The estimate for the background is equal to the number of observed events in the BG sample.

\subsection{\label{sec:mistag} Tagging matrix}

The tagging matrix is a parameterization of the heavy flavor rates as a function of the transverse energy of jets, the number of tracks associated to the jet and the number of primary vertices in the event.
Using the $b$-tagging algorithm described in Section~\ref{sec:EvSel}, we determine the above rates in a sample (4J) largely dominated by QCD multi-jet processes and selected from multi-jet data events with exactly four jets and passing the clean-up requirements described in Section~\ref{sec:EvSel}. 

\begin{figure}[!htbp]
\begin{center}
\includegraphics[width=\linewidth]{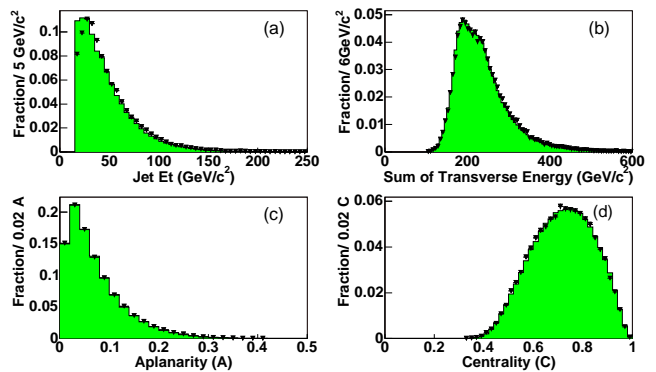}
\caption[Background validation in control region 1 for single tagged events]{Background validation in control region CR1 for single tagged events from the multi-jet data (dots) and from the background model (solid histogram). The distributions are normalized to the same area.}\label{fig:6-1}
\end{center}
\end{figure}

\begin{figure}[!htbp]
\begin{center}
\includegraphics[width=\linewidth]{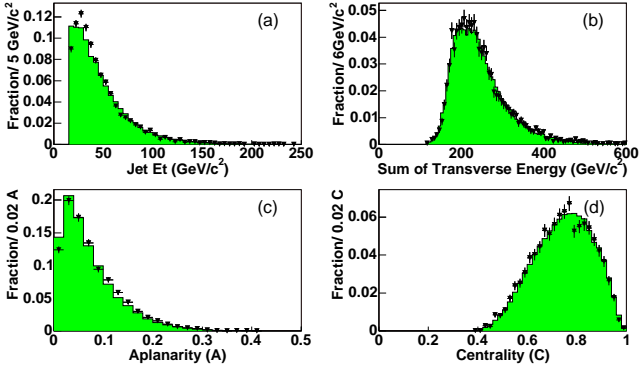}
\caption[Background validation in control region CR1 for double tagged events]{Background validation in control region CR1 for double tagged events from the multi-jet data (dots) and from the background model (solid histogram). The distributions are normalized to the same area.}\label{fig:6-2}
\end{center}
\end{figure}

We use a control region to check our assumption that the tagging rates from the 4-jet sample can be used to predict the tagging rates as a function of the variables used in the kinematical selection. 
This control region (CR1) contains events with exactly six jets and passing the clean-up cuts. 
The signal-to-background ratio in this region is about 1/250, estimated using same method as for the BG sample.
We compare the observed rates with the predicted rates based on the tagging matrix. 
Figure~\ref{fig:6-1} shows the comparison for events with exactly one secondary vertex tag, while Fig.~\ref{fig:6-2} shows the comparison in the sample with at least two secondary vertex tags. 
The variables chosen for this comparison are the transverse energies of jets, sum of the transverse energies of the six leading jets, aplanarity, and centrality as defined in Section~\ref{sec:EvSel}. 
The Kolmogorov-Smirnov probabilities for these comparisons in the single (double) tagged samples are: 0.0 (8.6E-5), 3E-11 (0.69), 0.99 (4.3E-3), and 0.12 (0.05), respectively.

Based on Fig.~\ref{fig:6-1}(a), the discrepancy between the observed rate and the predicted rate for jets with low transverse momentum may be an artifact of the binning of the tagging matrix. 
For transverse energies between 15~GeV and 40~GeV the tagging matrix uses the average rate, and therefore the rates for smaller intervals in this range might not be predicted well. 
Figures~\ref{fig:6-1}(a) and~\ref{fig:6-2}(a) support this by showing that, for this range of transverse energies, half of the data points are below and the other half is above the solid histogram representing our background model. 

The overall agreement between the observed and predicted rates is quite poor. In principle, a systematic uncertainty should be assigned to cover this discrepancies. However, the templates used in the mass measurement use the event probability based on matrix element information and they will be less affected by these inaccuracies. The reason for this is the fact that we use only a \tt matrix element. For background events the event probability (Eq.~\ref{eq:4-47})is flat as a function of the assumed top quark mass. The flatness of the event probability results in wide templates for the background sample and the systematic effects due to the mistag matrix will get smeared. In fact, the background templates in the control regions defined in Section~\ref{sec:bgproc} agree very well with the corresponding distributions based on the simulation of background events with $b\overline{b}$ + 4 light partons in the final state.

We conclude that the tagging matrix can be used to predict the background-like heavy flavor rates for events with same jet multiplicity as expected for the all-hadronic \tt events.
More details on the tagging matrix can be found in Ref.~\cite{ahxs}. 

\subsection{\label{sec:bgproc} Estimation of the background }

Based on the tagging matrix, a jet has a certain probability (rate) to be tagged as a heavy flavor jet depending on its transverse energy, number of tracks associated to it and number of vertices in the event. 
For a jet with a transverse energy between 15 GeV and 40 GeV and with ten associated tracks, this probability is (7.2 $\pm$ 0.5)$\%$. 
Using these probabilities we tag the jets as originating from a $b$ quark. 
This tagging procedure is repeated 20,000 times in the events of the BG sample producing about ten million tagged configurations which are interpreted as background events. 

A tagged configuration is an event from the BG sample where at least one of the six jets is tagged using the tagging matrix. 
Such kind of event can produce many tagged configurations which are unique if they have different tagged jets or a different number of tagged jets. 
We find 12888 unique single tagged configurations, and 26715 unique double tagged configurations.
Of these, 657 (or $\approx$ (5.1 $\pm$ 0.2)$\%$) single tagged configurations and 1180 (or $\approx$ (4.4 $\pm$ 0.1)$\%$) double tagged configurations pass the $minLKL$ cut.
We use these configurations, unique or duplicate, to form all relevant background distributions used for various checks and for the final measurement. 

The estimated number of background events is defined as the difference between the total number of events observed in the data sample and the expected number of \tt events based on the standard model expectation for \tt production cross section of 6.7~pb~\cite{Cacciari}.
This normalization applies to the top quark mass reconstruction procedure described in Section~\ref{sec:MassFit}, and for the validation of the background model described below.

We check various distributions of the background events modeled above against those from a sample of simulated events with $b\overline{b}$ + 4 light partons in the final state. 
This simulated sample is built using {\sc alpgen}~\cite{alpgen} for the event generation, {\sc pythia} for the parton showering, and the detector simulation as described in Section~\ref{sec:EvSel}. 
Given our event selection, other background sources are expected to have smaller contributions compared to the one from $b\overline{b}$ + 4 light partons and therefore affect less the relevant distributions. 

This check is performed in a control region (CR2) and in the signal region (SR) defined as follows. 
Region CR2 contains events that pass all our selection requirements without the $minLKL$ cut and has a signal-to-background ratio of about 1/6.
The signal region SR has events passing all selection criteria defined in Section~\ref{sec:EvSel}. 
Table~\ref{table:5-0} summarizes all the regions used in our background modeling procedure. 

\begin{table}[!htbp]
\begin{center}
\caption{Definition of the control regions used in the background modeling procedure. The selection requirements that differentiate them are defined in Section~\ref{sec:EvSel}.}\label{table:5-0} 
\begin{tabular*}{\linewidth}{c@{\extracolsep{\fill}}c@{\extracolsep{\fill}}c@{\extracolsep{\fill}}c@{\extracolsep{\fill}}c@{\extracolsep{\fill}}c@{\extracolsep{\fill}}c}
\hline
\hline
Region & Clean-up & $N_{jets}$ & Kinem. & $minLKL$ & $b$-tag & $N_{events}$\\
\hline
4J & yes & 4 & no & no & no & 2,242,512\\
BG & yes & 6 & yes & no & no & 2652\\
CR1 & yes & 6 & no & no & no & 380,676\\
CR2 & yes & 6 & yes & no & yes & 930\\
SR & yes & 6 & yes & yes & yes & 72\\
\hline
\hline
\end{tabular*}
\end{center}
\end{table}

Given that the BG sample used in our background model contains a small \tt content, we need to correct all the background distributions built from it. 
The relationship between a given uncorrected background distribution, $f_{B}$, and the corrected one, $f_{B}^{corr}$ is
\begin{equation}
\label{eq:5-1}f_{B}^{corr} = \frac{f_{B} - a_{S}f_{S}}{1-a_{S}}
\end{equation}
where $f_{S}$ is the corresponding distribution for \tt events and $a_{S}$ is the fraction of the uncorrected background sample due to \tt events. 
These quantities for \tt are determined from a sample of simulated \tt events where $M_{top}$ = 170~GeV/$c^{2}$ by randomly tagging the jets using the tagging matrix defined in Section~\ref{sec:mistag}. 
We choose the above value for the top quark mass based on the value of the world mass average~\cite{massavg} at the time of this analysis; in Section~\ref{sec:Syst} we determine a systematic uncertainty due to this choice.
The expression for $a_{S}$ in region CR2 is
\begin{equation}
\label{eq:5-2}a_{S}^{CR2} = \frac{N_{S}^{CR2}}{B^{CR2}+N_{S}^{CR2}}
\end{equation}
where $B^{CR2}$ is the background estimate in this region and $N_{S}^{CR2}$ is the number of \tt events estimated using the tagging matrix. 
The expression for $a_{S}$ in region SR is
\begin{equation}
\label{eq:5-3}a_{S}^{SR} = \frac{N_{S}^{CR2}\epsilon_{S}^{minLKL}}{B^{CR2}\epsilon_{B}^{minLKL}+N_{S}^{CR2}\epsilon_{S}^{minLKL}}
\end{equation}
where $\epsilon_{S}^{minLKL}$($\epsilon_{B}^{minLKL}$) is the efficiency of the $minLKL$ cut for \tt(background) in the CR2 region. 
The efficiency for background is determined using the ratio of the number of uniquely tagged configurations before the $minLKL$ cut (12,888 single tagged and 26,715 double tagged), and after the $minLKL$ cut respectively (657 single tagged and 1180 double tagged). 
Table~\ref{table:5-1} shows the estimated number of background events $B^{CR2}$ and the efficiency of the $minLKL$ cut for background $\epsilon_{B}^{minLKL}$ in region CR2. 
Tables~\ref{table:5-2} and~\ref{table:5-3} show the values for $\epsilon_{S}^{minLKL}$, $N_{S}^{CR2}$, and $a_{S}^{CR2}$ in region CR2 as well as the values of $a_{S}^{SR}$ for simulated \tt samples with different values on $M_{top}$.

\begin{table}[!htbp]
\begin{center}
\caption{The estimated number of background events $B^{CR2}$ and the efficiency of the $minLKL$ cut for background $\epsilon_{B}^{minLKL}$ in region CR2. The number of background events is the difference between the observed number of events and the expected number of \tt events assuming a \tt production cross section of 6.7~pb.}\label{table:5-1} 
\begin{tabular*}{\linewidth}{c@{\extracolsep{\fill}}c@{\extracolsep{\fill}}c}
\hline
\hline
Parameter & Single Tag & Double Tag \\
\hline
$B^{CR2}$ & 711 & 101\\
$\epsilon_{B}^{minLKL}$ & 0.051 & 0.044\\
\hline
\hline
\end{tabular*}
\end{center}
\end{table}

\begin{table}[!htbp]
\begin{center}
\caption{The number of \tt events, $N_{S}^{CR2}$, with one jet identified as $b$ jets using the tagging matrix; in region CR2, the acceptance of the $minLKL$ cut for \tt events, $\epsilon_{S}^{minLKL}$, and the values of the parameters $a_{S}^{CR2}$ (Eq.~\ref{eq:5-2}), and $a_{S}^{SR}$ (Eq.~\ref{eq:5-3}) for simulated \tt samples with different values on $M_{top}$.}\label{table:5-2} 
\begin{tabular*}{\linewidth}{c@{\extracolsep{\fill}}c@{\extracolsep{\fill}}c@{\extracolsep{\fill}}c@{\extracolsep{\fill}}c}
\hline
\hline
$M_{top}$ (GeV/$c^{2}$) & $N_{S}^{CR2}$ & $\epsilon_{S}^{minLKL}$ & $a_{S}^{CR2}$ & $a_{S}^{SR}$\\
\hline
160 & 29 & 0.21 & 0.039 & 0.146\\
170 & 30 & 0.20 & 0.040 & 0.144\\
175 & 28 & 0.19 & 0.038 & 0.130\\
180 & 28 & 0.18 & 0.038 & 0.124\\
\hline
\hline
\end{tabular*}
\end{center}
\end{table}

\begin{table}[!htbp]
\begin{center}
\caption{The number of \tt events, $N_{S}^{CR2}$, with at least two jets identified as $b$ jets using the tagging matrix; in region CR2, the acceptance of the $minLKL$ cut for \tt events, $\epsilon_{S}^{minLKL}$, and the values of the parameters $a_{S}^{CR2}$ (Eq.~\ref{eq:5-2}), and $a_{S}^{SR}$ (Eq.~\ref{eq:5-3}) for simulated \tt samples with different values on $M_{top}$.}\label{table:5-3} 
\begin{tabular*}{\linewidth}{c@{\extracolsep{\fill}}c@{\extracolsep{\fill}}c@{\extracolsep{\fill}}c@{\extracolsep{\fill}}c}
\hline
\hline
$M_{top}$ (GeV/$c^{2}$) & $N_{S}^{CR2}$ & $\epsilon_{S}^{minLKL}$ & $a_{S}^{CR2}$ & $a_{S}^{SR}$ \\
\hline

160 & 2 & 0.31 & 0.019 & 0.133\\
170 & 2 & 0.29 & 0.019 & 0.126\\
175 & 2 & 0.29 & 0.019 & 0.126\\
180 & 2 & 0.27 & 0.019 & 0.118\\
\hline
\hline
\end{tabular*}
\end{center}
\end{table}

The correction procedure uses by default the parameters as derived for $M_{top}$ = 170~GeV/$c^{2}$. 
In the determination of the systematic uncertainty due to this choice, we use the parameters corresponding to $M_{top}$ = 160~GeV/$c^{2}$, and $M_{top}$ = 180~GeV/$c^{2}$, respectively (see Section~\ref{sec:Syst}). 
The parameters obtained using $M_{top}$ = 175~GeV/$c^{2}$ are given for reference in Table~\ref{table:5-2} as that mass value corresponds to a \tt production cross section of 6.7~pb.

\begin{figure}[!htbp]
\begin{center}
\mbox{(a)}
\includegraphics[width=1.0\linewidth]{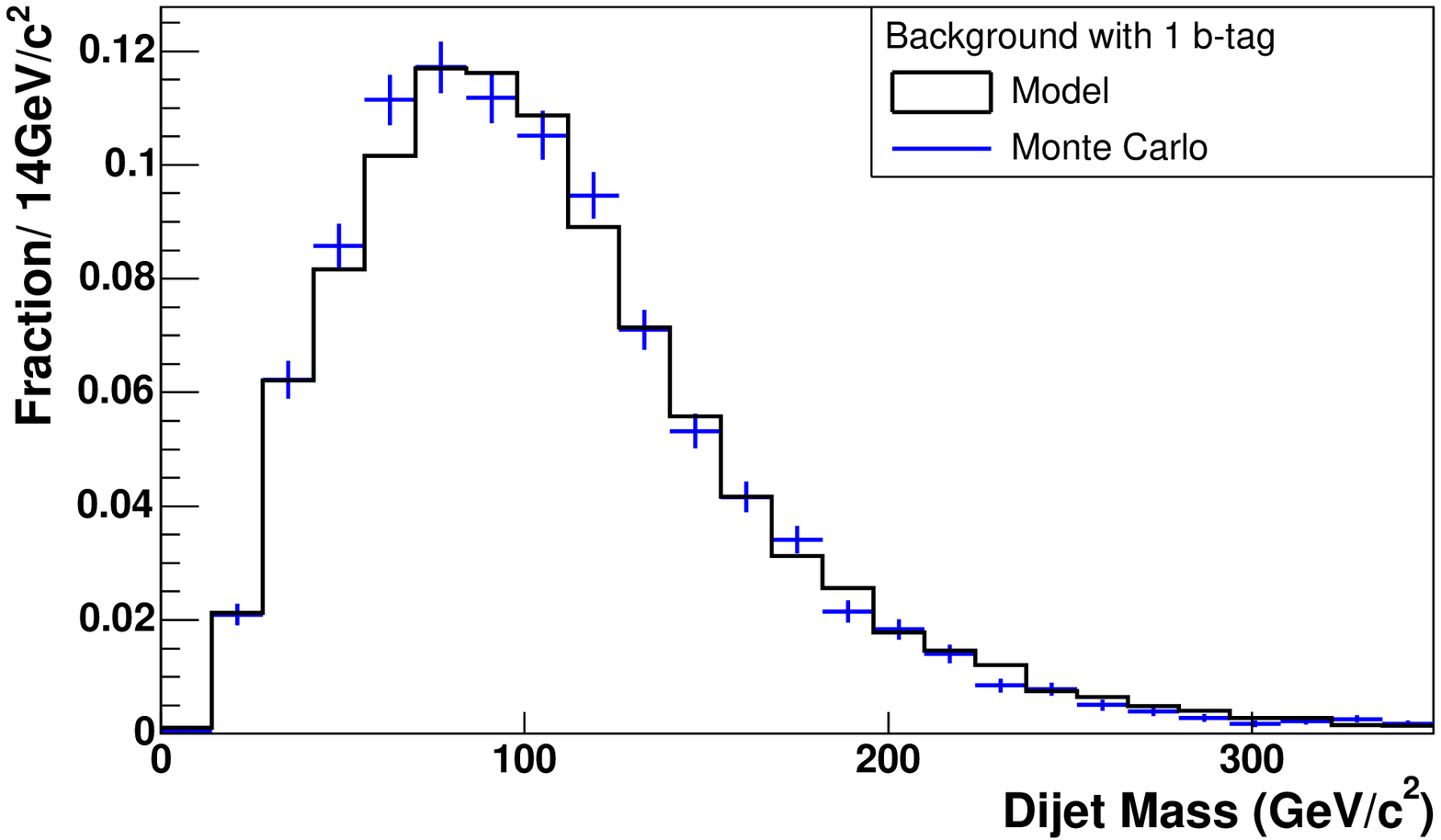}
\mbox{(b)}
\includegraphics[width=1.0\linewidth]{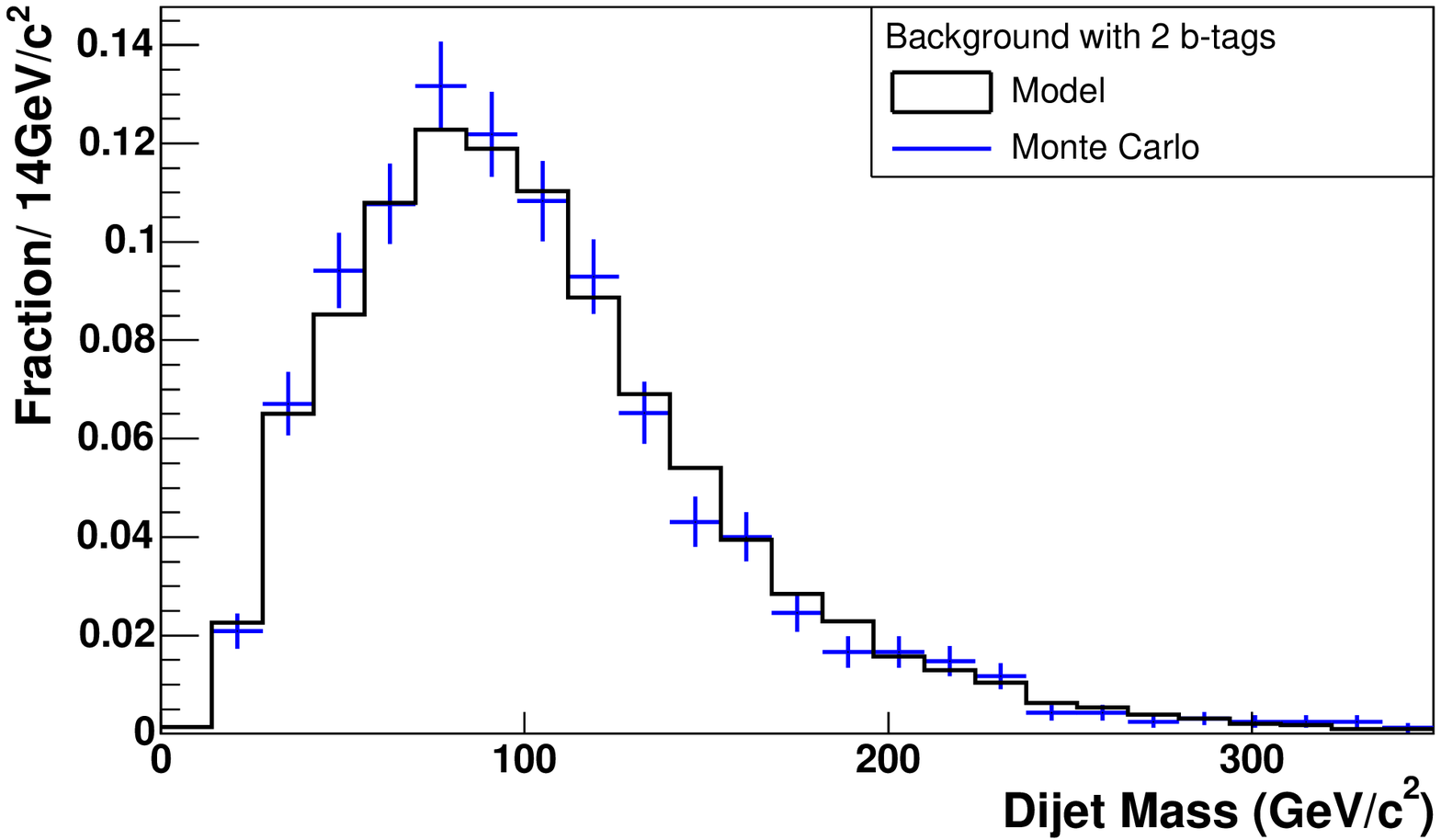}
\caption[Dijet invariant mass of untagged jets for background before the cut on the signal-like probability]{Invariant mass of pairs of untagged jets in control region CR2 for {\sc alpgen} $b\overline{b}$ + 4 light partons (cross), and for the background model (solid): (a) for single tagged events (Kolmogorov-Smirnov probability is $25\%$) and (b) for double tagged events (Kolmogorov-Smirnov probability is $43\%$).}\label{fig:6-4}
\end{center}
\end{figure}

\begin{figure}[!htbp]
\begin{center}
\mbox{(a)}
\includegraphics[width=1.0\linewidth]{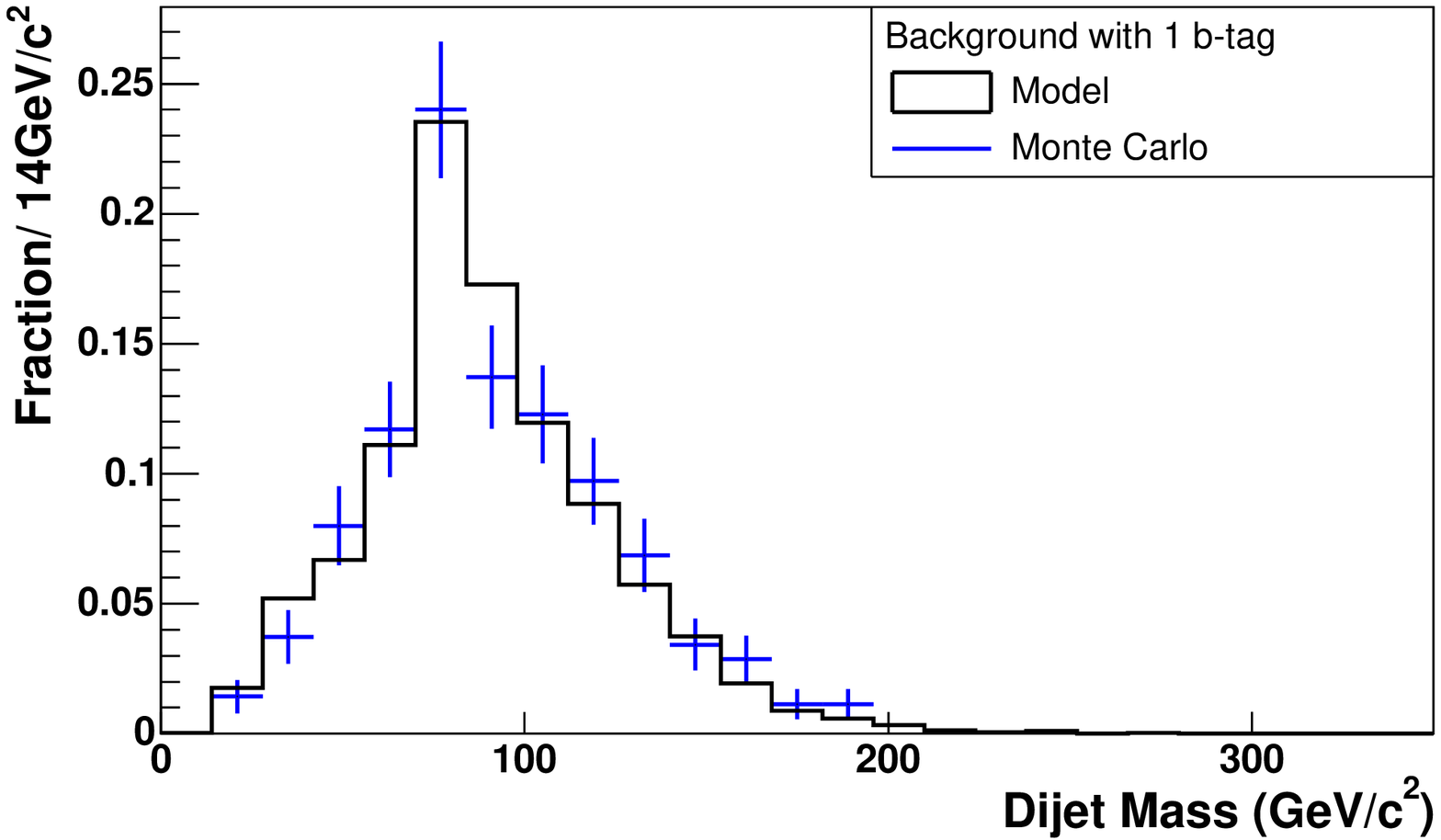}
\mbox{(b)}
\includegraphics[width=1.0\linewidth]{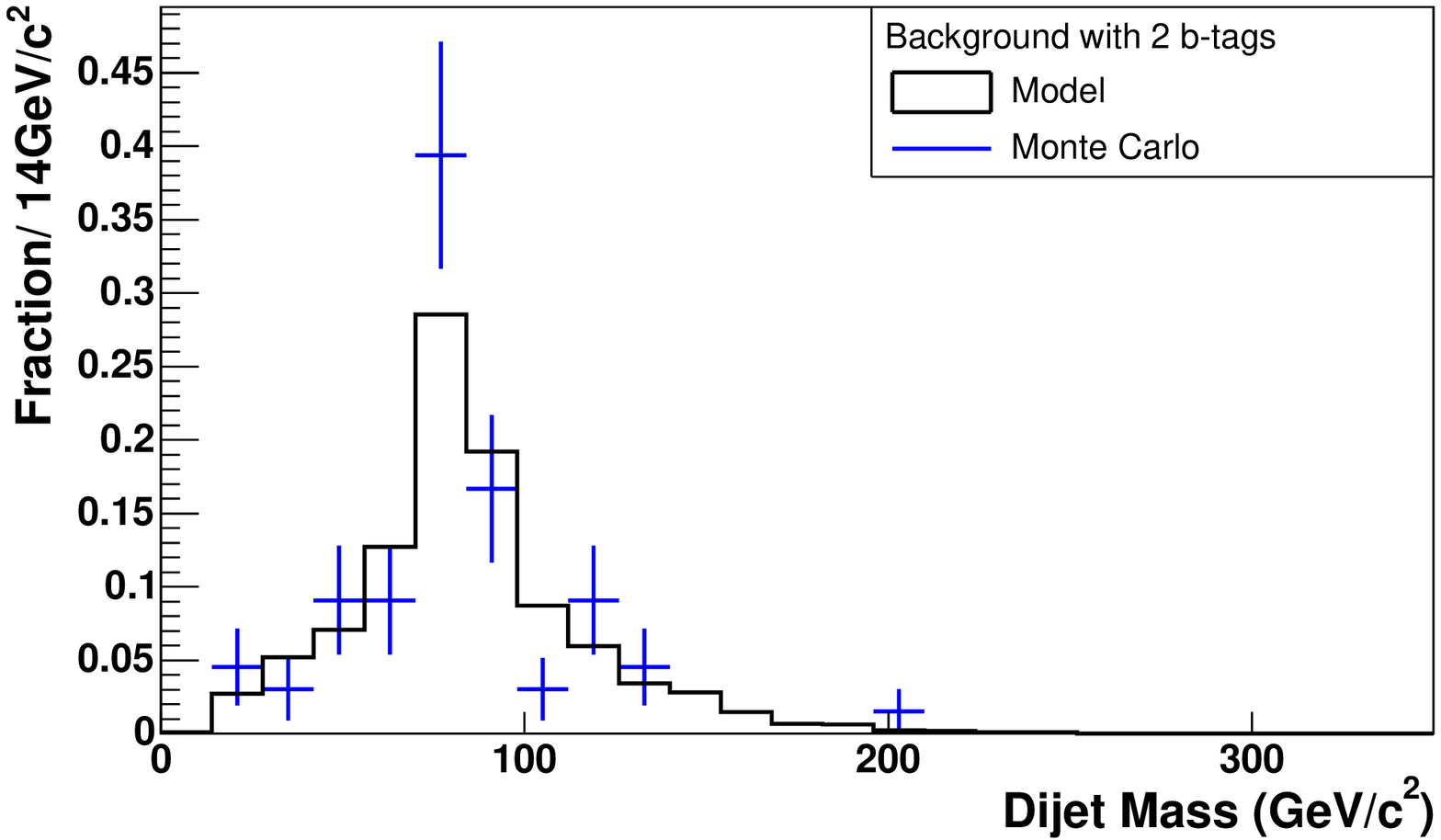}
\caption[Dijet invariant mass of untagged jets for background samples after the cut on the signal-like probability]{Invariant mass of pairs of untagged jets in signal region for {\sc alpgen} $b\overline{b}$ + 4 light partons (cross), and for the background model (solid): (a) for single tagged events (Kolmogorov-Smirnov probability is $90\%$), and (b) for double tagged events (Kolmogorov-Smirnov probability is $70\%$).}\label{fig:6-5}
\end{center}
\end{figure}

\begin{figure}[!htbp]
\begin{center}
\mbox{(a)}
\includegraphics[width=1.0\linewidth]{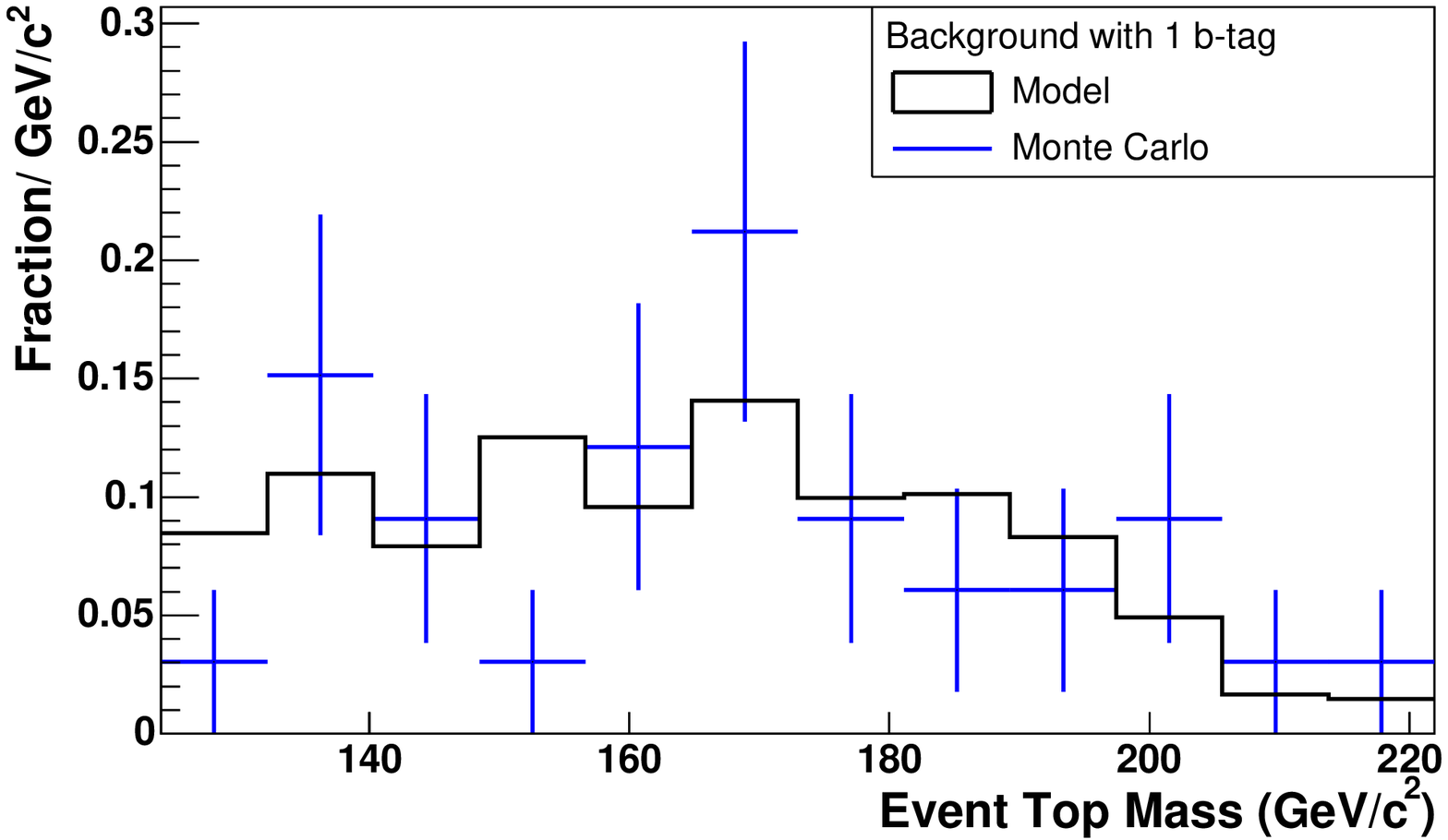}
\mbox{(b)}
\includegraphics[width=1.0\linewidth]{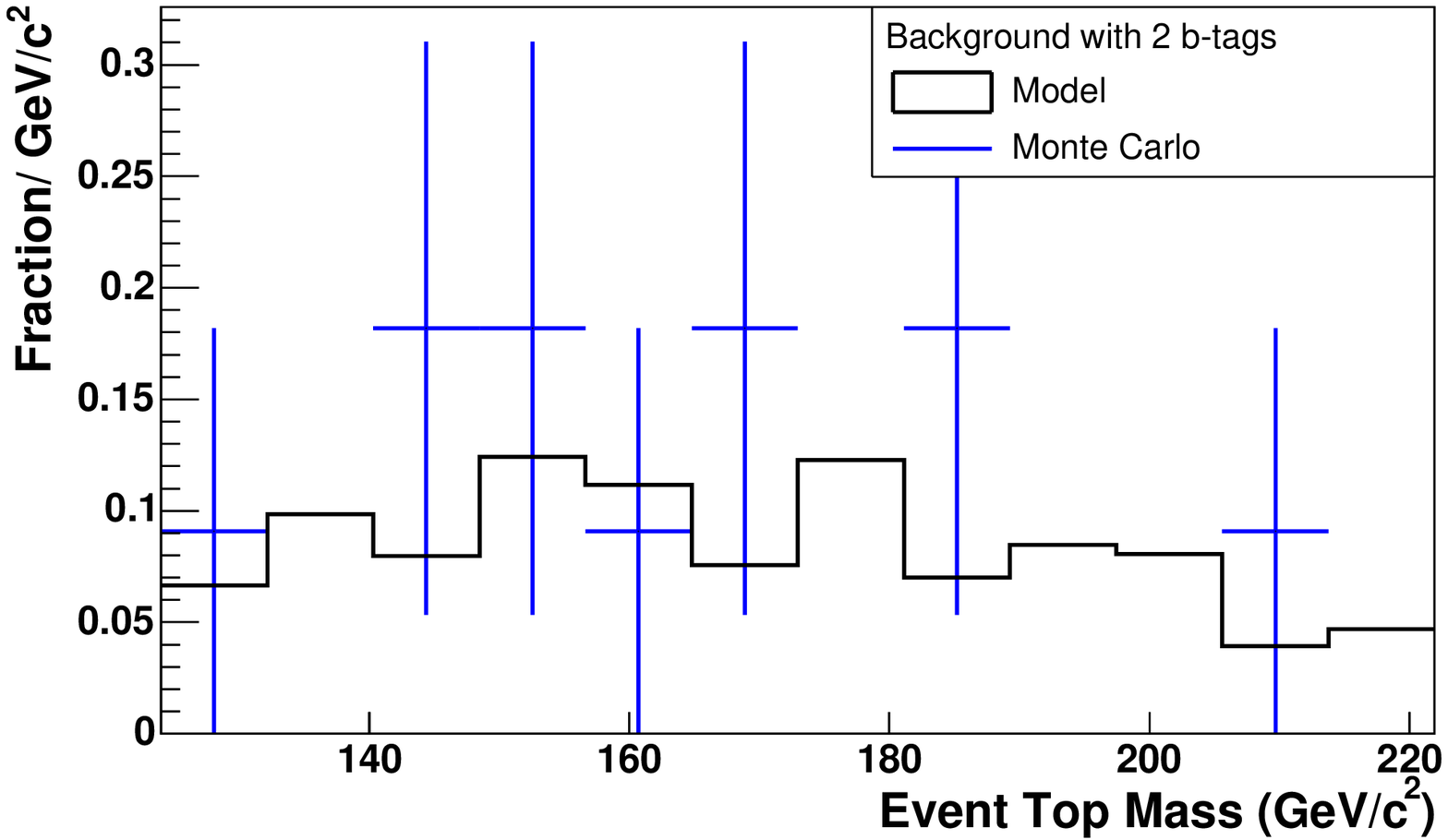}
\caption[Event by event most probable top quark mass distributions for background samples after the signal-like probability cut]{Event by event most probable top quark masses in the signal region for {\sc alpgen} $b\overline{b}$ + 4 light partons (cross), and for the background model (solid): (a) for single tagged events, and (b) for double tagged events.}\label{fig:6-6}
\end{center}
\end{figure}

Following this correction procedure, we compare shapes between our background model and the sample of simulated $b\overline{b}$ + 4 light partons described above. 
First, we do this comparison in region CR2 where we look at the invariant mass of all the untagged pairs of jets in the event (Fig.~\ref{fig:6-4}). 
The values of the Kolmogorov-Smirnov probabilities are 25$\%$ for the samples with single tagged events, and 43$\%$ for the samples with double tagged events.
For the signal region, we look at the invariant mass of all the untagged pairs of jets in the event (Fig.~\ref{fig:6-5}) and at the most probable per-event top quark mass (Fig.~\ref{fig:6-6}). 
These are variables of particular interest in this region as they will be used in the reconstruction of the top quark mass and for the  {\it in situ} calibration of the jet energy scale, as described in Section~\ref{sec:MassFit}. 
Based on the comparison from Fig.~\ref{fig:6-5}, the Kolmogorov-Smirnov probabilities are 90$\%$ for the single tagged events and 70$\%$ for the double tagged events. 

These comparisons show good agreement between our data-driven background model and a simulated sample of events containing the final state $b\overline{b}$ + 4 light partons, obtained using the {\sc alpgen} generator. 
In Section~\ref{sec:Syst} we evaluate the effect on the reconstructed top quark mass due to the limited statistics available in sample BG to construct the background model.

\section{\label{sec:MassFit} Top Quark Mass Estimation}

Our technique starts by modeling the data using a mixture of signal events obtained from \tt simulation and of background events obtained via our background model. 
The events are represented by two variables: the invariant mass of pairs of untagged jets and an event-by-event reconstructed top mass described below. 
These two variables are used to form distributions (templates), separately for \tt events and for background events. 
In the case of \tt events, the templates are parameterized as a function of the mass of the top quark and the jet energy scale (JES) variable (defined below).
For background no such dependences are expected since they contain no top quark and the jet energies used for the background modeling are taken from data. 
The measured values for the top quark mass and for the JES are determined using a likelihood technique described in Section~\ref{sec:lkl2D}. 

The largest contribution to the systematic uncertainty on the top quark mass is due to the uncertainty on the jet energy scale. 
To limit the impact of this systematic on the total uncertainty on the top quark mass, we use an {\it in situ} calibration of the jet energy scale via the $W$-boson mass. 
We measure a parameter JES that represents a shift in the jet energy scale from our default calibration as defined in Section~\ref{sec:EvSel}. 
This quantity is expressed in units of the total nominal jet energy scale uncertainty $\sigma_{c}$ that is derived following the default calibration. 
This uncertainty depends on the transverse energy, pseudorapidity, and the electromagnetic fraction of the jet energy. 
On average, the uncertainty is approximately equivalent to a 3$\%$ change in the jet energy scale for jets in \tt events. 
By definition, JES = 0 $\sigma_{c}$ represents our default jet energy scale; JES = 1 $\sigma_{c}$ corresponds to a shift in all jet energies by one standard deviation; and so on. 


The templates for \tt events are determined from samples of simulated \tt events with $M_{top}$ ranging from 150~GeV/$c^{2}$ to 200~GeV/$c^{2}$ in steps of 5~GeV/$c^{2}$.
We also include the sample where $M_{top}$ = 178~GeV/$c^{2}$ for a total of twelve different \tt simulated mass samples.
In addition to the variation of the top quark mass, for each value of $M_{top}$ we consider seven values for JES between -3 $\sigma_{c}$ and 3 $\sigma_{c}$, in steps of 1 $\sigma_{c}$. 
We use the events obtained from our background model to form the templates for the background.

\subsection{\label{sec:Templ} Definition and parameterization of the templates}

The first set of templates, called the top templates, is built using a variable ($m_{evt}^{top}$) determined using the matrix element technique. 
We call $m_{evt}^{top}$ the event-by-event reconstructed top quark mass, and it represents the mass value that maximizes the event probability defined in Section~\ref{sec:ME}.
We find the value of $m_{evt}^{top}$ by evaluating the event probability in the range 125~GeV/$c^{2}$ $\to$ 225~GeV/$c^{2}$. 
When building the templates, we drop the events for which the event probability is naturally maximized at mass values outside this range.
These events accumulate at the edges of the distribution making difficult the parameterization described below. 

\begin{figure}[!htbp]
\begin{center}
\includegraphics[width=\linewidth]{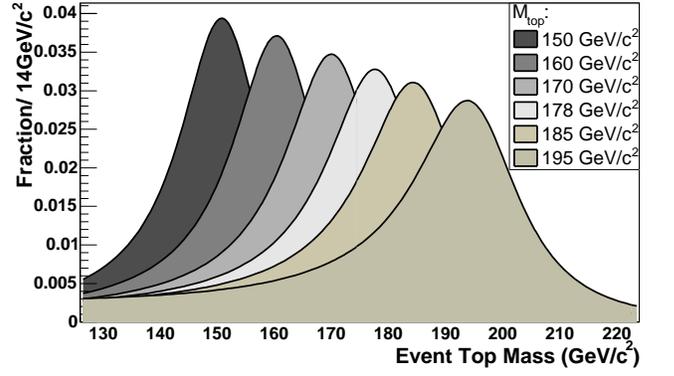}
\caption[Top templates for \tt events.]{The function fitting the top templates for \tt events at nominal JES and for various hypotheses of the top quark mass in the case of events with one tagged jet. A similar parameterization is obtained for events with at least two tagged jets.}\label{fig:7-1}
\end{center}
\end{figure}

For \tt events, the function $P_{s}^{top}(m_{evt}^{top}|M_{top},JES)$ used to describe the shape of these templates is a normalized product of a Breit-Wigner function and an exponential:
\begin{eqnarray}
P_{s}^{top}(m_{evt}^{top}|M_{top},JES) = \frac{\alpha_{0} \exp \left(-(m_{evt}^{top}-\alpha_{1})\alpha_{3} \right)}{N(M_{top},JES)} \nonumber \\
\label{eq:7-8} \times \frac{\alpha_{2} / 2\pi}{(m_{evt}^{top}-\alpha_{1})^{2} + \alpha_{2}^{2} / 4}
\end{eqnarray}
where the parameters $\alpha_{i}$ depend on $M_{top}$ and on JES. 
The normalization is set by $N(M_{top},JES)$ that has the following expression: 
\begin{eqnarray}
\label{eq:7-9} N(M_{top},JES) = \sum_{k=0}^{4} (p_{3k} + p_{3k+1}\cdot JES \nonumber \\
 + p_{3k+2}\cdot JES^{2})\cdot (M_{top})^{k}
\end{eqnarray}
The parameters $\alpha_{i}$ (Eq.~\ref{eq:7-8}) depend on $M_{top}$ and JES as follows:
\begin{equation}
\label{eq:7-10} \alpha_{i} = \left\{ 
\begin{array}{ll} 
p_{15} & i=0 \\ 
p_{3i+13} + p_{3i+14}\cdot M_{top} + p_{3i+15}\cdot JES & i=1,2,3
\end{array} 
\right.
\end{equation}
In Eqs.~\ref{eq:7-9} and~\ref{eq:7-10} the parameters $p_{i}$ are constants determined from the simultaneous fit of the top templates from all 84 \tt samples with the function $P_{s}^{top}(m_{evt}^{top}|M_{top},JES)$. 
Figure~\ref{fig:7-1} shows the function $P_{s}^{top}(m_{evt}^{top}|M_{top},JES)$ for JES = 0 $\sigma_{c}$ and various values of $M_{top}$ in the case of events with one tagged jet. 
A similar parameterization is obtained for events with at least two tagged jets.  

To determine how well the parameterization in Eq.~\ref{eq:7-8} describes the templates, we calculate the $\chi^{2}$ divided by the number of degrees of freedom, $N_{dof}$, as follows
\begin{equation}
\label{eq:7-11} \chi^{2}/N_{dof} = \frac{\sum_{m=1}^{12} \sum_{j=1}^{7} \sum_{bin=1}^{Nbins} \Big (\frac{h_{bin} - f_{bin}}{\sigma_{h_{bin}}} \Big)^{2}}{N_{dof}}
\end{equation}
where $h_{bin}$ is the bin content of the template histogram and $f_{bin}$ is the value of the function from Eq.~\ref{eq:7-8} at the center of the bin.
In  Eq.~\ref{eq:7-11}, the first two sums in the numerator are over the templates built from simulated \tt events for a given $M_{top}$ (12 values) and JES (7 values).
The third sum is over all the bins with more than 5 entries from each template. 
We obtain $\chi^{2}/N_{dof}=1554/1384=1.12$ for the sample with one secondary vertex tag and $\chi^{2}/N_{dof}=1469/1140=1.29$ for the sample with two secondary vertex tags corresponding to very small $\chi^{2}$ probabilities. 
From the values of the quantity $\chi^{2}/N_{dof}$, we conclude that the parameterization of the top templates is not very accurate, and we expect some bias in the reconstruction of mass and JES. 
The procedure for bias removal is described in Section~\ref{sec:Check}.

\begin{figure}[!htbp]
\begin{center}
\mbox{(a)}
\includegraphics[width=1.0\linewidth]{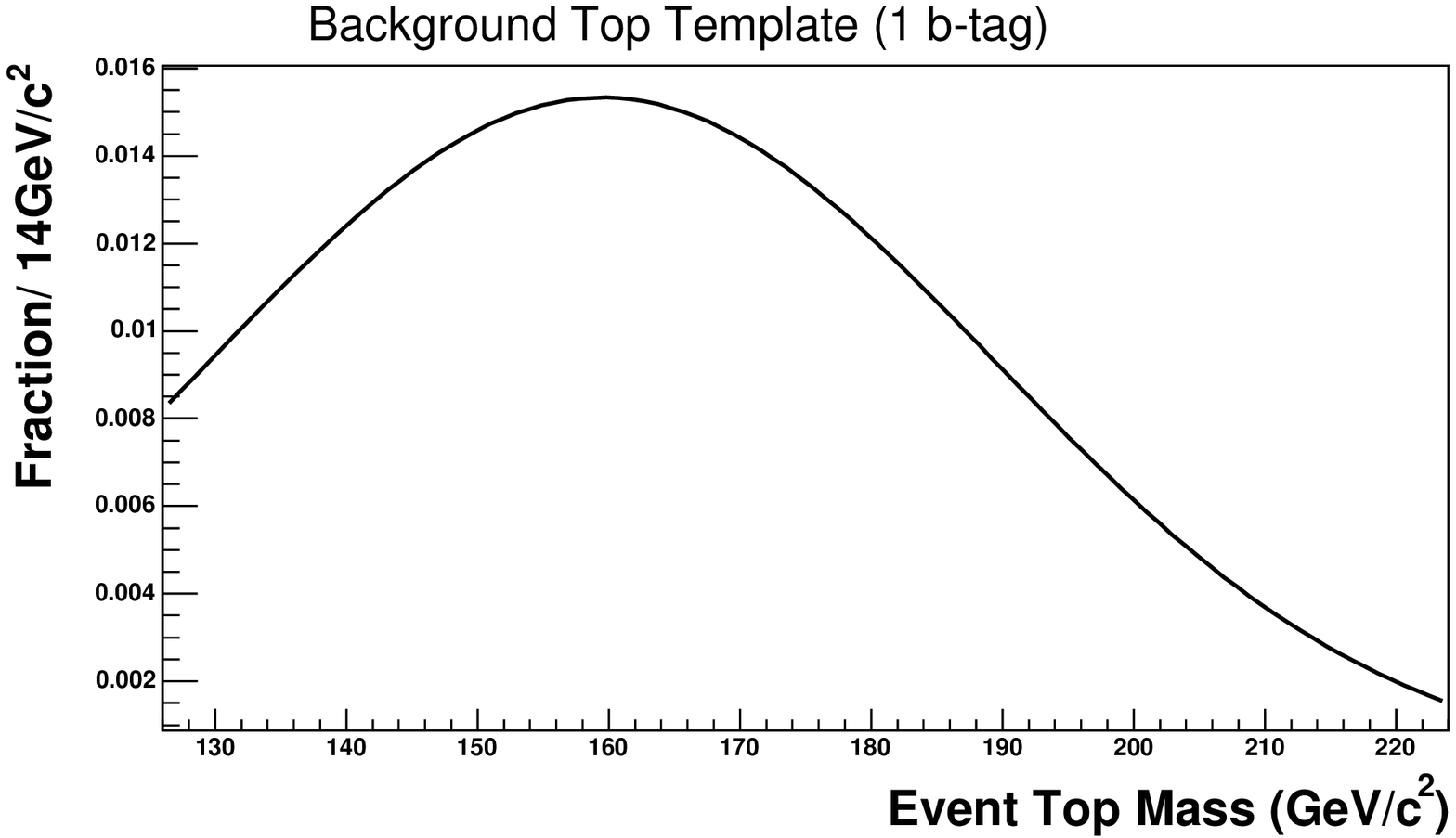}
\mbox{(b)}
\includegraphics[width=1.0\linewidth]{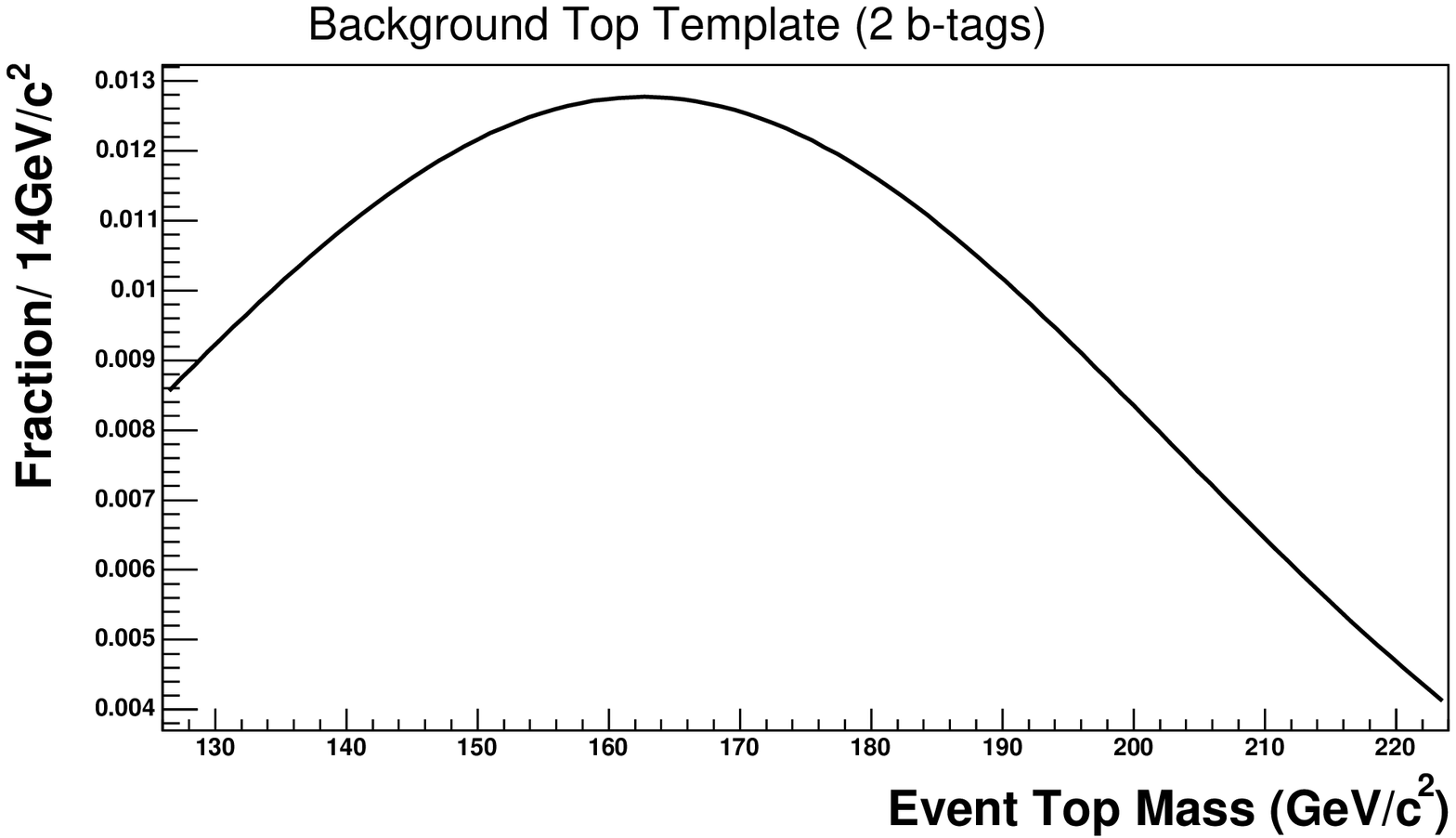}
\caption[Top templates for background events]{Top templates for (a) single tagged background events and for (b) double tagged background events.}\label{fig:7-2}
\end{center}
\end{figure}

The top templates for background events are built using the matrix element in the same way as for \tt events. 
The shape of the background template is fitted to a normalized Gaussian. 
Figure~\ref{fig:7-2} shows separately the resulting parameterized curves of background templates for single and double tagged background events. 

\begin{figure}[!htbp]
\begin{center}
\includegraphics[width=\linewidth]{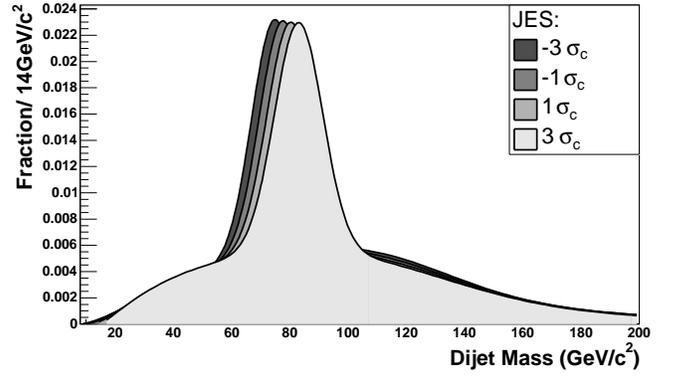}
\caption[Dijet mass templates for \tt events.]{The function fitting the dijet mass templates for \tt events with $M_{top}$ = 170~GeV/$c^{2}$ and various values of JES in the case of events with one tagged jet. A similar parameterization is obtained for events with at least two tagged jets.}\label{fig:7-3}
\end{center}
\end{figure}

The second set of templates, the dijet mass templates, are formed by considering the invariant mass $m_{evt}^{W}$ of all possible pairs of untagged jets in the sample. 
This variable is correlated to the mass of the $W$ boson, and plays a central role in the {\it in situ} calibration of the jet energy scale. 
For \tt events the function $P_{s}^{W}(m_{evt}^{W}|M_{top},JES)$ used to fit the dijet mass templates is a normalized sum of two Gaussians and a Gamma function:
\begin{eqnarray}
P_{s}^{W}(m_{evt}^{W}|M_{top},JES) = \frac{1}{N'(M_{top},JES)} \nonumber \\
\times \left[ \frac{\beta_{6}\beta_{7}  \exp \big(-\beta_{7} (m_{evt}{W}-\beta_{8}) \big) }{\Gamma (1+\beta_{9}) } \cdot (m_{evt}{W} - \beta_{8})^{\beta_{9}} \right. \nonumber \\
+ \left. \frac{\beta_{0}}{\beta_{2}\sqrt 2\pi} \exp \left(-\frac{(m_{evt}^{W}-\beta_{1})^{2}}{2\beta_{2}^{2}} \right) \right.  \nonumber \\
\label{eq:7-12} + \left . \frac {\beta_{3}}{\beta_{5}\sqrt 2\pi} \exp \left(-\frac{(m_{evt}^{W}-\beta_{4})^{2}}{2 \beta_{5}^{2}}\right) \right]
\end{eqnarray}
where the parameters $\beta_{i}$ depend on $M_{top}$ and on JES.
The normalization is set by $N'(M_{top},JES)$ that has the following expression:
\begin{eqnarray}
\label{eq:7-13} N'(M_{top},JES) = \sum_{k=0}^{1} (q_{3k} + q_{3k+1}\cdot JES \nonumber \\
+ q_{3k+2}\cdot JES^{2})\cdot (M_{top})^{k}
\end{eqnarray}
The parameters $\beta_{i}$ depend on $M_{top}$ and JES as follows:
\begin{equation}
\label{eq:7-14} \beta_{i} = q_{3i+6} + q_{3i+7}\cdot M_{top} + q_{3i+8}\cdot JES , i=0,9
\end{equation}
In Eqs.~\ref{eq:7-13} and~\ref{eq:7-14} the parameters $q_{i}$ are constants determined from the simultaneous fit of the top templates from all 84 \tt samples with the function $P_{s}^{W}(m_{evt}^{W}|M_{top},JES)$. 
Figure~\ref{fig:7-3} shows the function $P_{s}^{W}(m_{evt}^{W}|M_{top},JES)$ for $M_{top}$ = 170~GeV$c^{2}$ and various values of JES in the case of events with one tagged jet. 
A similar parameterization is obtained for events with at least two tagged jets. 

As in the case of top templates, we calculate (Eq.~\ref{eq:7-11}) the quantity $\chi^{2}/N_{dof}$ to describe the performance of the parameterization of the dijet mass templates. 
We obtain $\chi^{2}/N_{dof}=3551/2636=1.35$ for the sample with one secondary vertex tag and $\chi^{2}/N_{dof}=2972/2524=1.18$ for the sample with at least two secondary vertex tags. 
From the values of the quantity $\chi^{2}/N_{dof}$ we reach the same conclusion as in the case of the parameterization of top templates: the parameterization of the dijet mass templates is not very accurate and some bias is expected when the top mass and JES are reconstructed. 

\begin{figure}[!htbp]
\begin{center}
\mbox{(a)}
\includegraphics[width=1.0\linewidth]{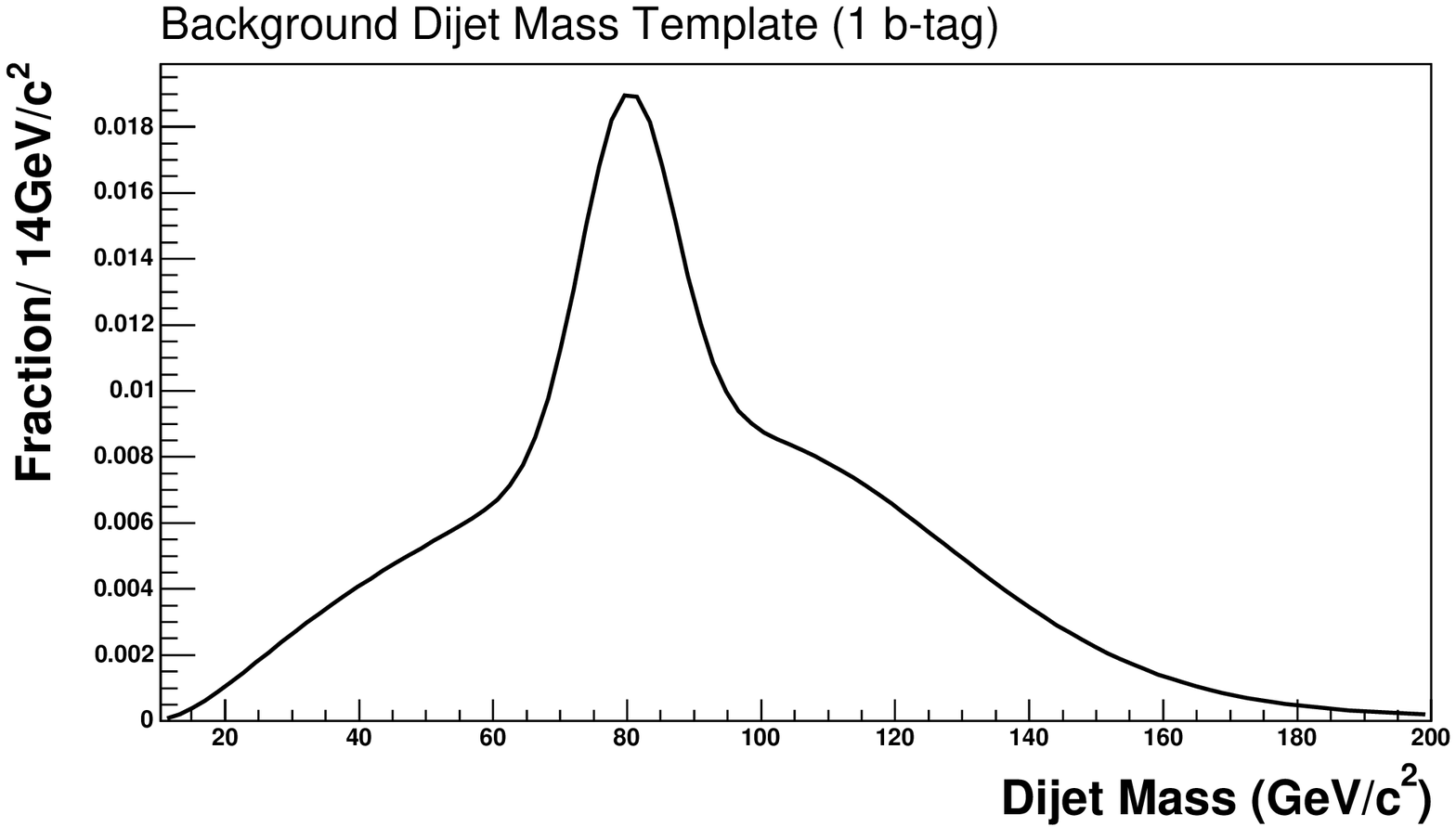}
\mbox{(b)}
\includegraphics[width=1.0\linewidth]{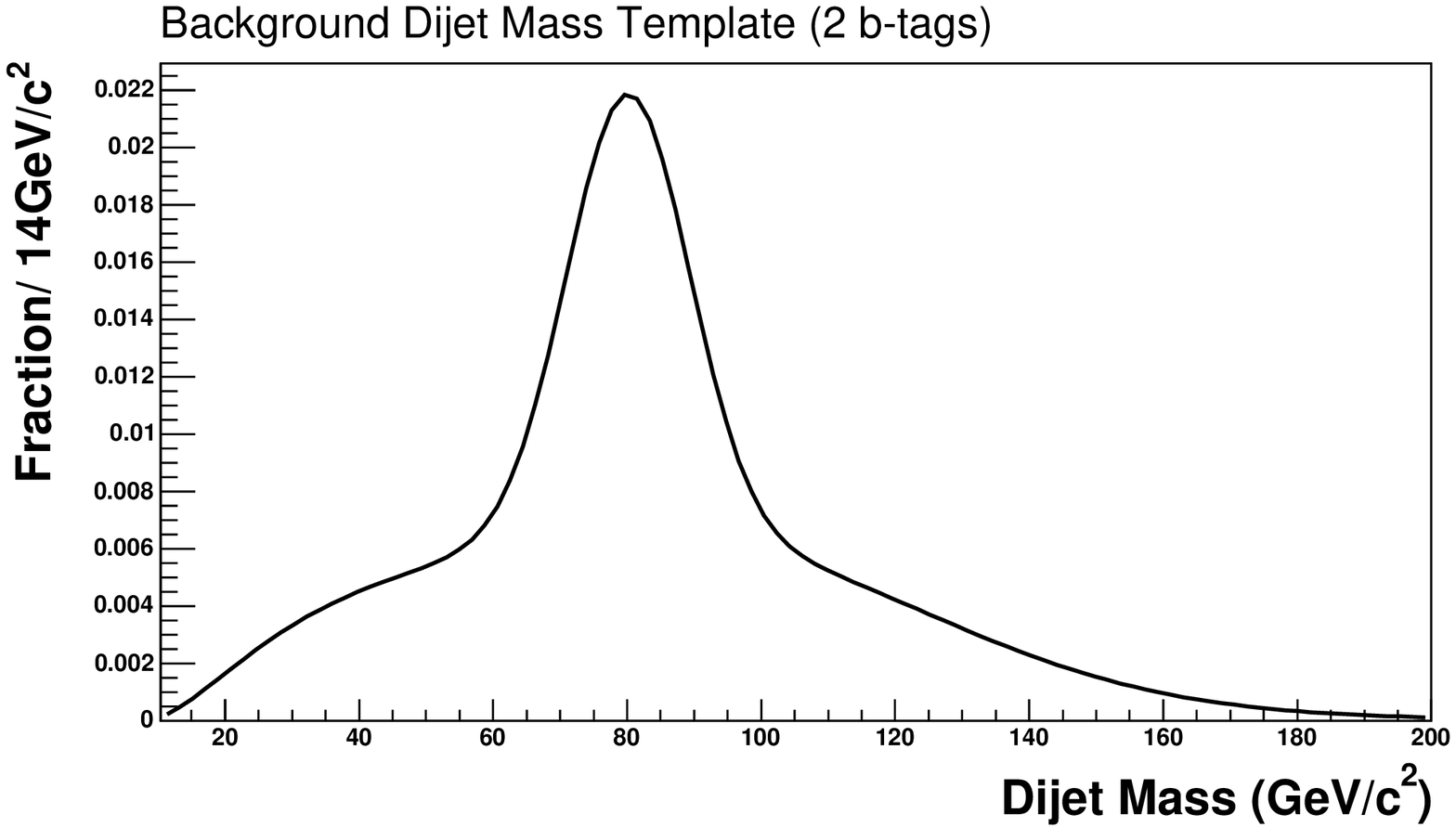}
\caption[Dijet mass templates for background events]{Dijet mass templates for (a) single tagged background events and for (b) double tagged background events.}\label{fig:7-4}
\end{center}
\end{figure}

The dijet mass template for background is built in the same way as for the \tt templates. 
The background template is fitted to a normalized sum of two Gaussians and a Gamma function. 
This combination of functions provided the best fit of the dijet mass shapes. 
Figure~\ref{fig:7-4} shows separately the resulting parameterized curves of background templates for single and double tagged background events. 

\subsection{\label{sec:lkl2D} Likelihood definition}

The mass of the top quark and the value of JES are determined by maximizing a likelihood function built using the two sets of templates described in Section~\ref{sec:Templ}. 
Assuming that the data sample is the sum of $n_{s}$ \tt events and $n_{b}$ background events, we can calculate the likelihood function connected to a generic template $P^{f}$ as
\begin{eqnarray}
\label{eq:7-3} \mathcal{L}_{f}(M_{top},JES) = \prod_{evt=1}^{N_{evt}^{tot}} \left( \frac{n_{s} \cdot P_{s}^{f}(x_{evt}|M_{top},JES)}{n_{s}+n_{b}} \right. \nonumber \\
+ \left. \frac{n_{b} \cdot P_{b}^{f}(x_{evt})}{n_{s}+n_{b}} \right)
\end{eqnarray}
where index $f$ can either be $top$ when the variable $x_{evt}$ represents the event-by-event reconstructed top mass, or $W$ for the invariant mass of pairs of light flavor jets.

The number of \tt events, $n_{s}$, is constrained to the expected number of \tt events, $n_{s}^{exp}$, via a Gaussian
\begin{equation}
\label{eq:7-6} \mathcal{L}_{n_{s}} = \exp \left( - \frac{(n_{s}-n_{s}^{exp})^{2}}{2 \sigma_{n_{s}^{exp}}} \right)
\end{equation}
with mean equal to $n_{s}^{exp}$ and width equal to $\sigma_{n_{s}^{exp}}$, the uncertainty on the expected number of \tt events. 

The expected numbers of signal events, $n_{s}^{exp}$, are 13 for the single tagged sample and 14 for the double tagged sample corresponding to a theoretical cross section of $6.7_{-0.9}^{+0.7}$ pb~\cite{Cacciari} and an integrated luminosity of 943 pb$^{-1}$. 
The value of the theoretical cross section assumes a top quark mass of 175~GeV/$c^{2}$. 
The values for $\sigma_{n_{s}^{exp}}$ are 3.7 for the single tagged sample and 3.9 for the double tagged sample, which take into account both statistical effects (assuming a Poisson distribution) on $n_{s}^{exp}$ and systematic ones based on the uncertainty on the theoretical cross section. 

The sum of \tt and background events, $n_{s}+n_{b}$, is constrained to the total number of observed events in the data, $N_{evt}^{tot}$, via a Poisson probability with a mean equal to $N_{evt}^{tot}$ 
\begin{equation}
\label{eq:7-5} \mathcal{L}_{nev} = \frac {(N_{evt}^{tot})^{n_{s}+n_{b}} \exp (-N_{evt}^{tot})}{(n_{s}+n_{b})!}
\end{equation}

Multiplying the terms expressing the constraints on the number of events and the likelihood functions for each template, we obtain separate likelihood functions for events with one tag and for events with at least two tags:
\begin{equation}
\label{eq:7-2} \mathcal{L}_{n-tag} = \mathcal{L}_{top} \cdot \mathcal{L}_{W} \cdot \mathcal{L}_{nev} \cdot \mathcal{L}_{n_{s}}
\end{equation}

As described in Section~\ref{sec:EvSel}, the jet energy scale JES can be determined from independent detector calibrations. 
We include this knowledge in the likelihood in the form of a Gaussian constraint on our variable JES. 
This Gaussian has a mean equal to the expectation on JES from the independent calibration, JES$^{exp}$ = 0 $\sigma_{c}$, and a width equal to 1 $\sigma_{c}$ which is the uncertainty on this expectation.
\begin{equation}
\label{eq:7-7} \mathcal{L}_{JES} = \exp \left( - \frac{(JES-JES^{exp})^{2}}{2} \right)
\end{equation}

The term expressing the constraint on the JES variable is multiplied together with the likelihood function for each heavy flavor sample to obtain the final likelihood function used to reconstruct the top quark mass shown in Eq.~\ref{eq:7-1}.
\begin{equation}
\label{eq:7-1} \mathcal{L} = \mathcal{L}_{1tag} \cdot \mathcal{L}_{2tag} \cdot \mathcal{L}_{JES}
\end{equation}

Following the maximization of the likelihood function shown in Eq.~\ref{eq:7-1} we will obtain six numbers: the reconstructed top quark mass $M_{t}$, the reconstructed JES variable $JES_{out}$, and the number of events with different number of tags for \tt, $n_{1,2}^{S}$, and for background, $n_{1,2}^{B}$. 
The statistical uncertainties on these numbers, $\delta M_{t}$, $\delta JES_{out}$, $\delta n_{1,2}^{S}$, and $\delta n_{1,2}^{B}$ are obtained from the points where the log-likelihood changes by 0.5.

\subsection{\label{sec:Check} Calibration of the method}

Using samples of simulated \tt events and the background sample built based on the model presented in Section~\ref{sec:Bckd}, we form simulated experiments for a series of JES and $M_{top}$ input values. 
We then verify that the reconstructed values of the top quark mass and JES obtained following the maximization of the likelihood function (Section~\ref{sec:lkl2D}) are in agreement with the input values. 
The simulated experiments are a mixture of \tt events and background events reflecting the expected sample composition of the data.
In each simulated experiment, the number of \tt events is drawn from a Poisson distribution of mean equal to the expected number of \tt events passing the selection, as determined from simulation (Table~\ref{table:x6-1}).
The number of background events is also drawn from a Poisson distribution with a mean equal to the difference between the observed number of events (see Section~\ref{sec:EvSel}, Table~\ref{table:5-4}) and the expected number of \tt events.

\begin{table}[!htbp]
\begin{center}
\caption{Number of events for samples of simulated \tt events with $M_{top}$ ranging between 150~GeV/$c^{2}$ and 200~GeV/$c^{2}$. The numbers correspond to a integrated luminosity of 943 pb$^{-1}$, after all selection requirements are made. The observed number of events is also shown.}\label{table:x6-1} 
\begin{tabular*}{\linewidth}{c@{\extracolsep{\fill}}c@{\extracolsep{\fill}}c}
\hline
\hline
$M_{top}$ (GeV/$c^{2}$) & Single Tag & Double Tag \\
\hline
150 & 18 & 14 \\
155 & 17 & 15 \\
160 & 16 & 14 \\
165 & 16 & 14 \\
170 & 15 & 14 \\
175 & 13 & 14 \\
178 & 14 & 14 \\
180 & 12 & 13 \\
185 & 11 & 11 \\
190 & 9 & 11 \\
195 & 9 & 10 \\
200 & 7 & 8 \\
\hline
Total Observed & 48 & 24 \\
\hline
\hline
\end{tabular*}
\end{center}
\end{table}

In order to reduce the statistical uncertainties on potential biases in mass or JES reconstruction, about 10,000 simulated experiments are performed. 
Due to the finite size of simulated \tt event samples and background sample the simulated experiments share events between them. 
These overlaps result in correlations between the results of the mass and JES reconstructions from each simulated experiment. 
These correlations are taken into account following the study found in Ref.~\cite{Barlow}. 
The typical value for the correlation between any two simulated experiments is 6$\%$.

The variables extracted from each simulated experiment are: the values of mass, $M_{t}^{PE}$, and JES, $JES_{out}^{PE}$ that maximize the likelihood defined in section~\ref{sec:lkl2D}; the statistical uncertainties on the above variables, $\delta M_{t}^{PE}$ and $\delta JES_{out}^{PE}$ and the pulls for these variables as defined by
\begin{eqnarray}
\label{eq:8-4} Pull_{mass} = \frac{M_{t}^{PE}-M_{top}}{\delta M_{t}^{PE}} \nonumber \\ 
Pull_{JES} = \frac{JES_{out}^{PE}-JES_{true}}{\delta JES_{out}^{PE}}
\end{eqnarray}
where JES$_{true}$ is the value of JES used in the simulation. 

The distribution of the top quark masses $M_{t}^{PE}$ reconstructed in each simulated experiment is fitted to a Gaussian. 
The mean of this Gaussian is interpreted as the reconstructed top quark mass of the sample, $M_{t}$, while the width of the Gaussian represents the expected statistical uncertainty on it, $\delta M_{t}$.
We apply the same procedure to determine the reconstructed value of JES, $JES_{out}$, and its expected statistical uncertainty, $\delta JES_{out}$.

\begin{figure}[!htbp]
\begin{center}
\includegraphics[width=\linewidth]{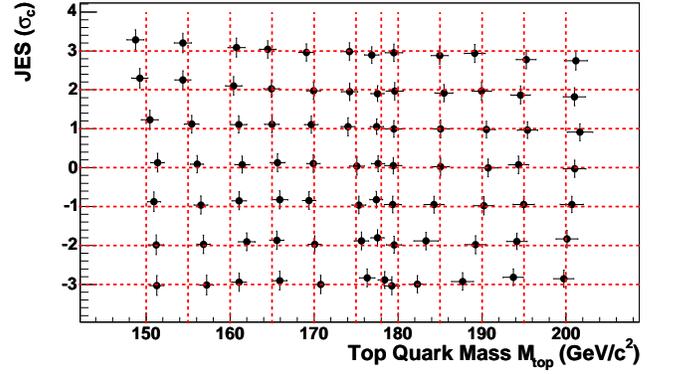}
\caption[Raw reconstruction in the JES versus Top Mass plane]{JES versus Top Quark Mass plane. The points represent the reconstructed JES, $JES_{out}$, and top quark mass $M_{t}$ and have attached their corresponding statistical uncertainties, $\delta JES_{out}$ and $\delta M_{t}$. The vertical dashed lines correspond to the true values of the mass, while the horizontal lines correspond to the true values of JES. For a perfect reconstruction the points should sit right at the intersection of the dashed lines.}\label{fig:8-1}
\end{center}
\end{figure}

\begin{figure}[!htbp]
\begin{center}
\includegraphics[width=\linewidth]{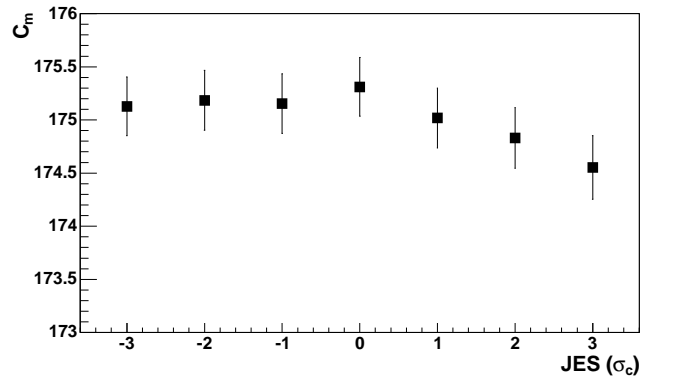}
\caption{Distribution of parameter $C_{m}$ (Eq.~\ref{eq:8-7}) as a function of JES.}\label{fig:cm}
\end{center}
\end{figure}

\begin{figure}[!htbp]
\begin{center}
\includegraphics[width=\linewidth]{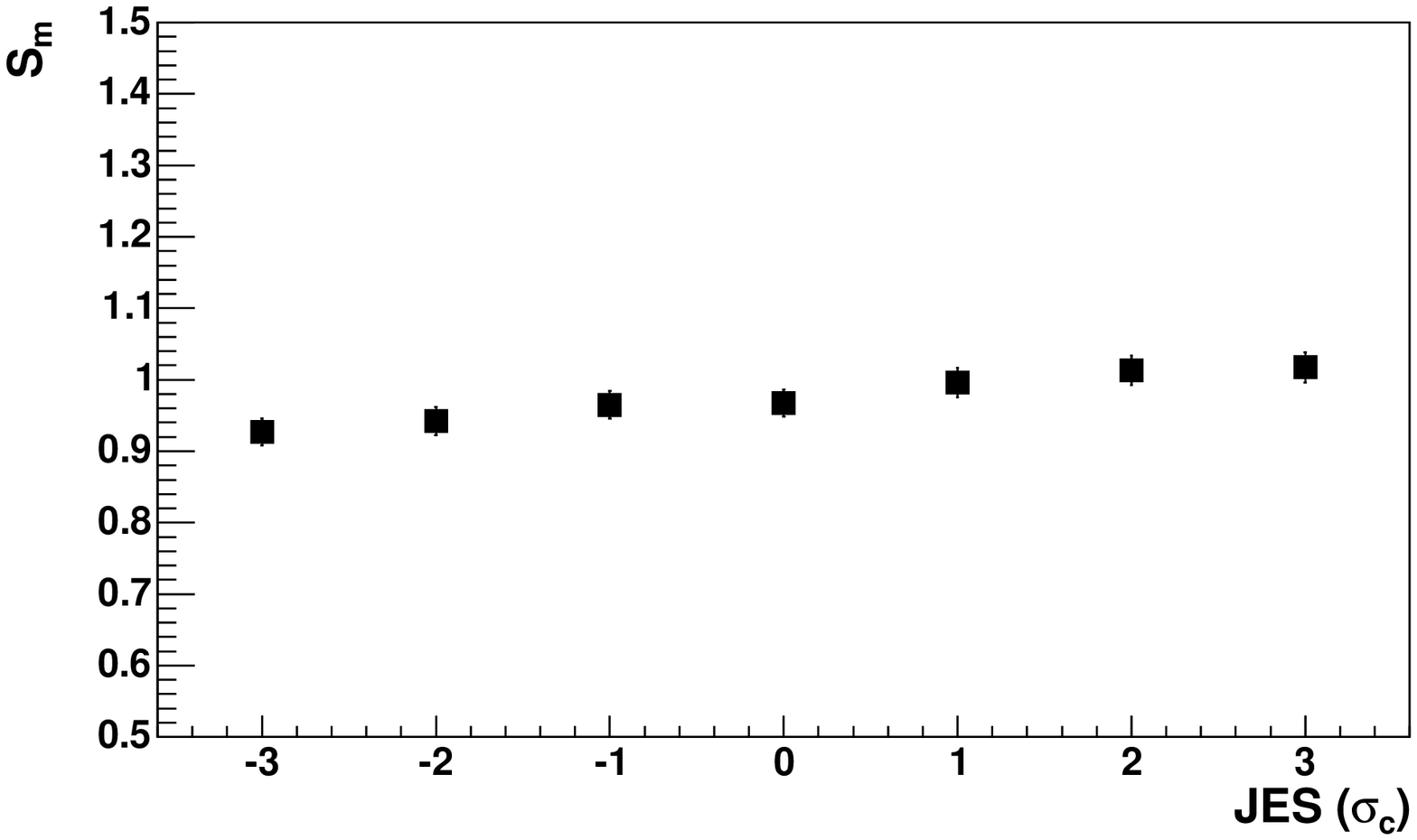}
\caption{Distribution of parameter $S_{m}$ (Eq.~\ref{eq:8-7}) as a function of JES.}\label{fig:sm}
\end{center}
\end{figure}

\begin{figure}[!htbp]
\begin{center}
\includegraphics[width=\linewidth]{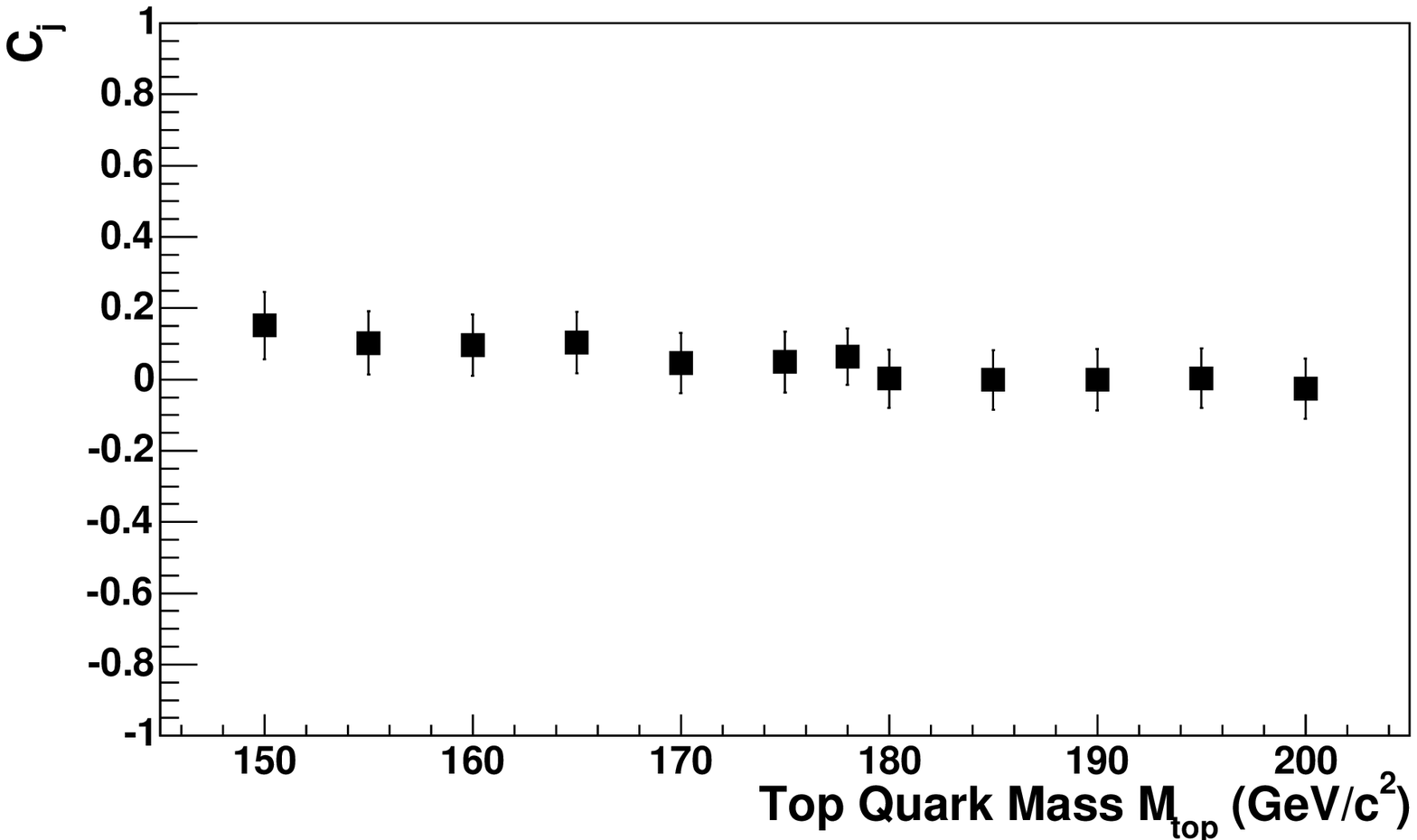}
\caption{Distribution of parameter $C_{j}$ (Eq.~\ref{eq:8-7}) as a function of $M_{top}$.}\label{fig:cj}
\end{center}
\end{figure}

\begin{figure}[!htbp]
\begin{center}
\includegraphics[width=\linewidth]{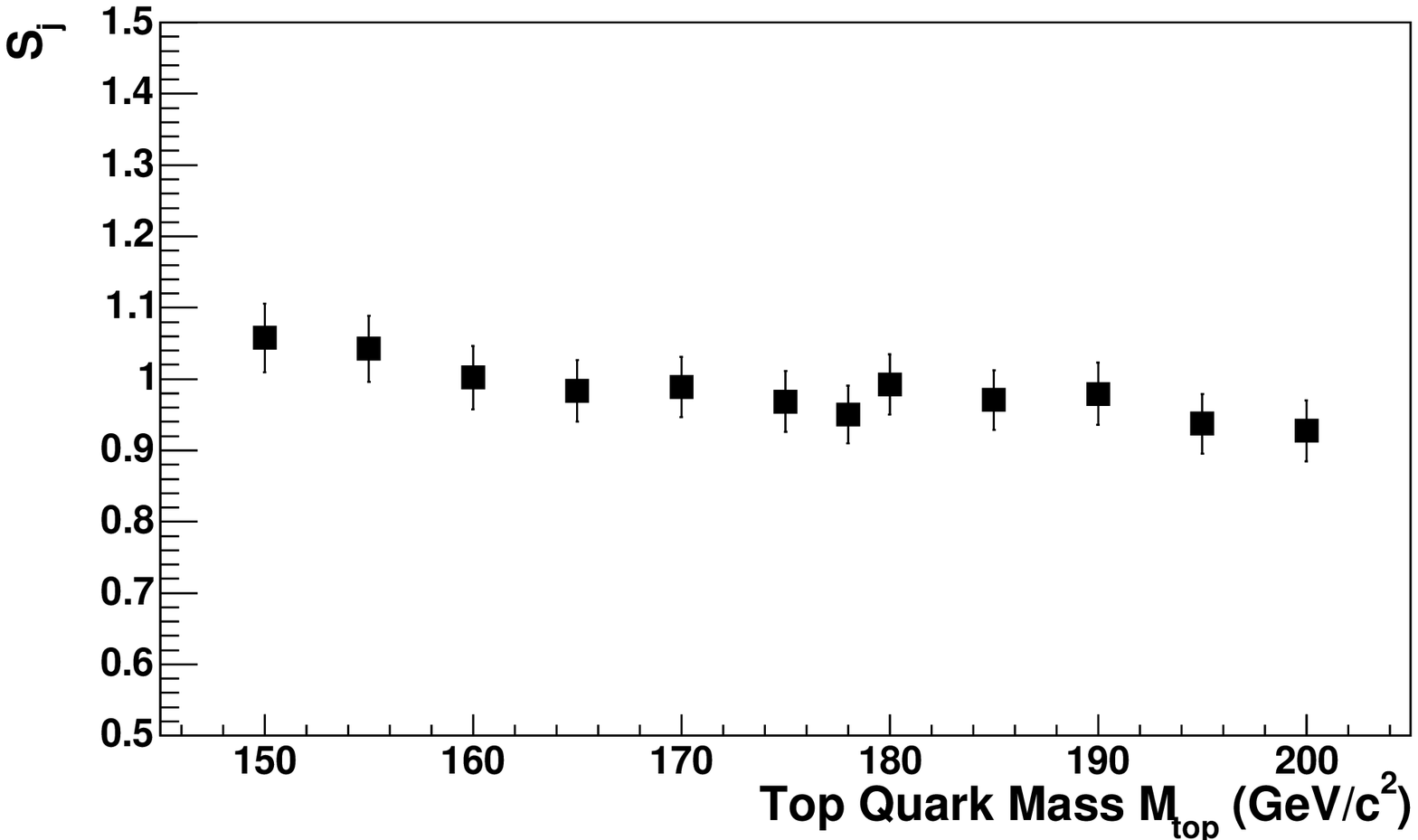}
\caption{Distribution of parameter $S_{j}$ (Eq.~\ref{eq:8-7}) as a function of $M_{top}$.}\label{fig:sj}
\end{center}
\end{figure}

Figure~\ref{fig:8-1} shows the reconstructed JES and the reconstructed top mass represented by the points, versus the true JES and true top mass represented by the grid. 
Ideally the points should match the grid crossings, but there is a slight bias which has to be removed. 
The bias is removed in the mass-JES plane by solving the system in Eq.~\ref{eq:8-6}
\begin{eqnarray}
M_{t} & = & C_{m} + S_{m} \cdot (M_{top}-175) \nonumber \\
\label{eq:8-6} JES_{out} & = & C_{j} + S_{j} \cdot JES_{true}
\end{eqnarray}
for $M_{top}$ and JES$_{true}$. 
The parameters $C_{m}$, $C_{j}$, $S_{m}$, and $S_{j}$ have the form 
\begin{equation}
\label{eq:8-7} \begin{array}{lll}
C_{m} & = & a_{1} + a_{2} \cdot JES_{true} \\
S_{m} & = & a_{3} + a_{4} \cdot JES_{true} \\
C_{j} & = & b_{1} + b_{2} \cdot M_{top} \\
S_{j} & = & b_{3} + b_{4} \cdot M_{top} 
\end{array}
\end{equation}
where the parameters $\{a_{i}\}$ and $\{b_{i}\}$ from Eq.~\ref{eq:8-7} are listed in Table~\ref{table:8-2}. 
They are determined from a linear fit of the distributions of $C_{m}$ and $S_{m}$  versus JES$_{true}$ (Figs.~\ref{fig:cm} and~\ref{fig:sm}), and of $C_{j}$ and $S_{j}$  versus $M_{top}$, respectively (Figs.~\ref{fig:cj} and~\ref{fig:sj}).

\begin{table}[!htbp]
\begin{center}
\caption[Linearity check of the $M_{top}$ and JES reconstruction]{Values of the parameters describing best the linear dependence on the true JES and on the true $M_{top}$, of the intercept and slope of the $M_{top}$ calibration curve and of the JES calibration curve respectively.}\label{table:8-2}
\begin{tabular*}{\linewidth}{c@{\extracolsep{\fill}}r@{.\extracolsep{0pt}}l@{\extracolsep{\fill}}r@{.\extracolsep{0pt}}l}
\hline
\hline
Parameter & \multicolumn{2}{c}{Value} & \multicolumn{2}{c}{Uncertainty}\\
\hline
$a_{1}$ & 175&0 & 0&1 \\
$a_{2}$ & -0&09 & 0&05 \\
$a_{3}$ & 0&975 & 0&008 \\
$a_{4}$ & 0&016 & 0&004 \\
\hline
$b_{1}$ & 0&6 & 0&3 \\
$b_{2}$ & -0&003 & 0&002 \\
$b_{3}$ & 1&35 & 0&15 \\
$b_{4}$ & -0&0021 & 0&0008 \\
\hline
\hline
\end{tabular*}
\end{center}
\end{table}

The uncertainties $\delta M_{t}$ and $\delta JES_{out}$ on the reconstructed values $M_{t}$ and JES$_{out}$ are also affected by the bias in the reconstruction technique and we need to correct them as well. 
By differentiating Eq.~\ref{eq:8-6} with respect to $M_{top}$ and JES$_{true}$, we obtain another system of equations to be solved for the corrected uncertainties, $\delta M_{t}^{corr}$ and $\delta JES_{out}^{corr}$. 
\begin{eqnarray}
\delta M_{t} & = & X_{m} \cdot \delta JES_{true} + Y_{m} \cdot \delta M_{t}^{corr} \nonumber \\
\label{eq:8-9} \delta JES_{out} & = & X_{j} \cdot \delta M_{t}^{corr} + Y_{j} \cdot \delta JES_{out}^{corr}
\end{eqnarray}

The parameters $X_{m}$, $X_{j}$, $Y_{m}$, and $Y_{j}$ from Eq.~\ref{eq:8-9} depend on $M_{top}$ and JES$_{true}$ as shown in Eq.~\ref{eq:8-10}. 
Solving Eq.~\ref{eq:8-9} provides the best estimate of the uncertainties on $M_{t}$ and on JES$_{out}$.
\begin{equation}
\label{eq:8-10} \begin{array}{lll}
X_{m} & = & a_{2} + a_{4} \cdot (M_{top}-175) \\
Y_{m} & = & a_{3} + a_{4} \cdot JES_{true} \\
X_{j} & = & b_{2} + b_{4} \cdot JES_{true} \\
Y_{j} & = & b_{3} + b_{4} \cdot M_{top} \\
\end{array}
\end{equation}

\begin{figure}[!htbp]
\begin{center}
\includegraphics[width=1.0\linewidth]{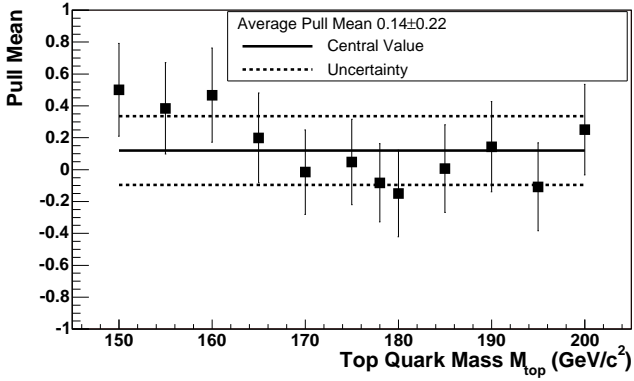}
\caption{Pull means versus $M_{top}$ in the case of the reconstruction of top quark mass in samples with JES$_{true}$ = 0 $\sigma_{c}$. The continuous line represents the average pull mean and the dashed lines show the uncertainty on it.}\label{fig:8-8}
\end{center}
\end{figure}

\begin{figure}[!htbp]
\begin{center}
\includegraphics[width=1.0\linewidth]{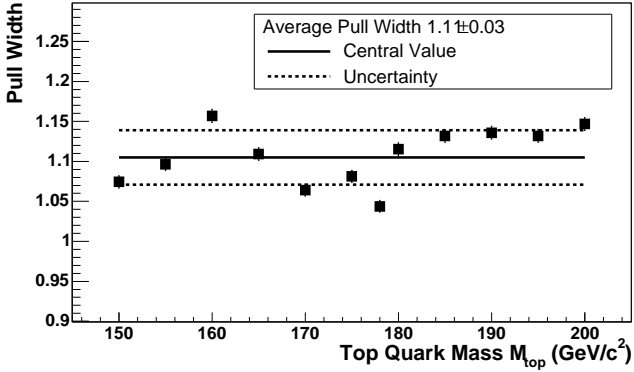}
\caption{Pull widths versus $M_{top}$ in the case of the reconstruction of top quark mass in samples with JES$_{true}$ = 0 $\sigma_{c}$. The continuous line represents the average pull width and the dashed lines show the uncertainty on it.}\label{fig:8-9}
\end{center}
\end{figure}

Following the procedure for removing the bias in the mass reconstruction, the distribution of pull means extracted using simulated experiments (Fig.~\ref{fig:8-8}) validates our bias correction as, on average, the pull mean is estimated to be consistent with zero within the uncertainty. 
The width of the pull distribution is used to determine the corrections on the statistical uncertainties $\delta M_{t}^{corr}$ due to non-Gaussian behavior of the likelihood function (Eq.~\ref{eq:7-1}). 
Figure~\ref{fig:8-9} shows the mass pull widths versus top quark mass $M_{top}$. 
In these plots the JES$_{true}$ of the \tt samples is 0 $\sigma_{c}$. 
Similar pulls are obtained from \tt samples with different values of JES$_{true}$. 
Based on these figures, it is estimated that the uncertainty on $M_{t}$ has to be increased by 11$\%$. 

\begin{figure}[!htbp]
\begin{center}
\includegraphics[width=1.0\linewidth]{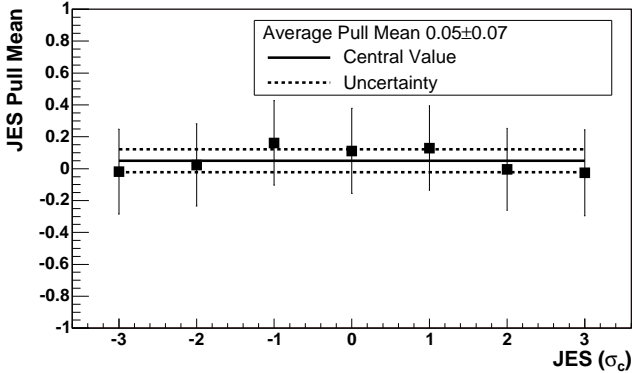}
\caption{Pull means versus $JES_{true}$ in the case of the reconstruction of JES in samples with $M_{top}$ = 170~GeV/$c^{2}$. The continuous line represents the average pull mean and the dashed lines show the uncertainty on it.}\label{fig:8-12}
\end{center}
\end{figure}

\begin{figure}[!htbp]
\begin{center}
\includegraphics[width=1.0\linewidth]{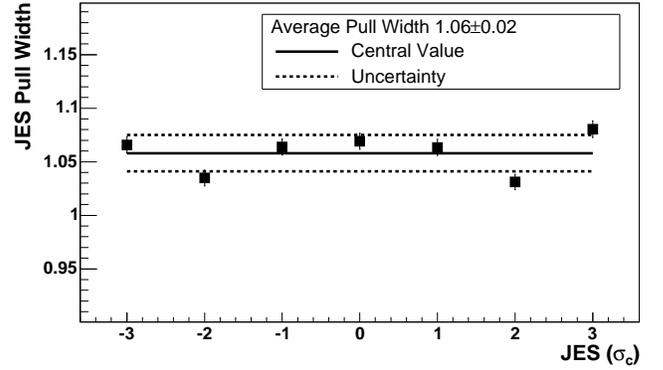}
\caption{Pull widths versus versus $JES_{true}$ in the case of the reconstruction of JES in samples with $M_{top}$ = 170~GeV/$c^{2}$. The continuous line represents the average pull width and the dashed lines show the uncertainty on it.}\label{fig:8-13}
\end{center}
\end{figure}

For the reconstruction of JES, Fig.~\ref{fig:8-12} shows the pull means versus JES$_{true}$, while Fig.~\ref{fig:8-13} shows the pull widths versus JES$_{true}$. 
In both plots, $M_{top}$ = 170~GeV/$c^{2}$. 
Similar pulls are obtained from \tt samples with different values of $M_{top}$. 
Regarding the bias correction, we reach the same conclusion as in the case of the mass reconstruction that, on average, the pull mean is estimated to be consistent with zero within the uncertainties. 
Based on Fig.~\ref{fig:8-13}, it is estimated that the uncertainty on the JES$_{out}$ has to be increased by 6$\%$.

In order to further establish the robustness of the technique, the mass and JES are measured in samples for which the true values are unknown to the authors of this paper. 
To validate the mass reconstruction we utilize five such blind samples: three generated with {\sc Herwig} and two with {\sc Pythia}. 
The value of JES in these samples corresponds to 0, the nominal jet energy scale. 
The reconstructed top quark mass in each of these samples is the most probable value obtained from 10,000 simulated experiments. 
Each simulated experiment is formed combining the \tt events in the blind samples and the background events from the background model such that on average the total number of events is equal to the observed value (see Table~\ref{table:x6-1}). 
The size of the \tt content is 15 single tagged events and 14 double tagged events. 

Following the mass reconstruction technique and the calibration described in this paper, the differences between the true top quark mass values and the reconstructed ones are: -0.2, 0.3, 0.6, -0.7, and 1.1~GeV/$c^{2}$. 
The statistical uncertainty on these numbers is 0.8~GeV/$c^{2}$. 
The first two numbers correspond to the {\sc Pythia} samples. 
To validate the JES reconstruction, another five blind samples are used for which the jet energy scale is modified. 
The generator used here is {\sc Herwig} and the value of the top quark mass is 170~GeV/$c^{2}$. 
The differences between the true JES values and the reconstructed ones are: 0.1, 0.3, 0.0, 0.1, and -0.1 $\sigma_{c}$. 
The statistical uncertainty on these numbers is 0.4 $\sigma_{c}$. 

In conclusion, both the mass and JES reconstructed values are compatible with true ones within the statistical uncertainties. 
This additional check gave us confidence that the method described here can be reliably applied on the data to reconstruct JES and the top quark mass.

\subsection{\label{sec:Stat} Expected statistical uncertainty}

\begin{figure}[!htbp]
\begin{center}
\includegraphics[width=\linewidth]{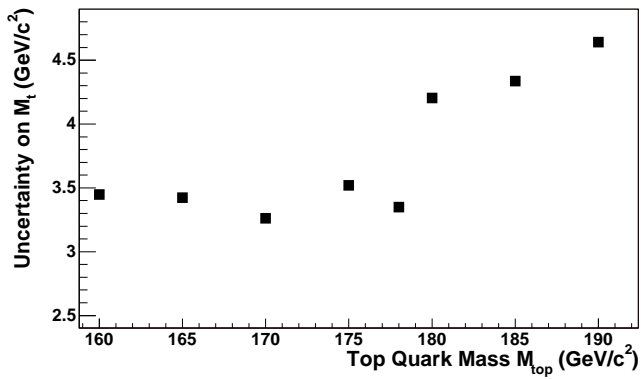}
\end{center}
\caption[Expected uncertainty on top mass versus input top mass]{Expected uncertainty on top quark mass, $\delta M_{t}^{corr}$, versus $M_{top}$, for samples with JES$_{true}$ = 0 $\sigma_{c}$. This uncertainty includes the uncertainty due to statistical effects and the systematic uncertainty due to jet energy scale.}\label{fig:8-23}
\end{figure}

\begin{figure}[!htbp]
\begin{center}
\includegraphics[width=\linewidth]{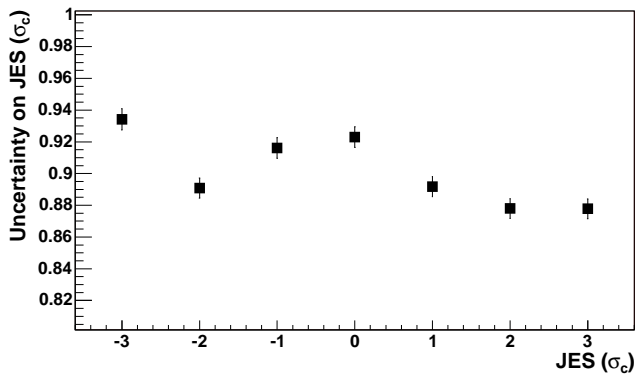}
\end{center}
\caption[Expected uncertainty on JES versus input JES]{Expected uncertainty on JES, $\delta JES_{out}$, versus JES$_{true}$, for samples with $M_{top}$ = 170~GeV/$c^{2}$.}\label{fig:8-24}
\end{figure}

In Fig.~\ref{fig:8-23} we show the expected uncertainty on top quark mass, $\delta M_{t}^{corr}$, versus $M_{top}$, for samples with JES$_{true}$ = 0 $\sigma_{c}$. 
Since the expected number of \tt events depends on the $M_{top}$, the uncertainty $\delta M_{t}^{corr}$ depends on it too. 
Figure~\ref{fig:8-24} shows the expected uncertainty on JES, $\delta JES_{out}^{corr}$, versus JES$_{true}$, for samples with $M_{top}$ = 170~GeV/$c^{2}$. 
The uncertainties in Fig.~\ref{fig:8-23} and~\ref{fig:8-24} are corrected for bias, but not for pull widths (non-Gaussian effects). 

The expected uncertainties shown in Fig.~\ref{fig:8-23} contain both the statistical uncertainty on the top quark mass and the uncertainty due to jet energy scale. 
In order to disentangle the statistical uncertainty on $M_{t}$ from the one due to jet energy scale, we reconstruct the top quark mass by maximizing the likelihood for a fixed value of JES. 
Following this reconstruction for $M_{top}$ = 170~GeV/$c^{2}$ and JES$_{true}$ = 0 $\sigma_{c}$, the uncertainty on the top quark mass is 2.5~GeV/$c^{2}$. 
In comparison, when JES is not fixed the expected uncertainty (Fig.~\ref{fig:8-23}) on $M_{t}$ is 3.2~GeV/$c^{2}$. 
Subtracting these two numbers in quadrature we estimate that the systematic uncertainty on $ M_{t}$ due to jet energy scale is 2.0~GeV/$c^{2}$. 

We can determine the systematic uncertainty on $M_{t}$ due to the jet energy scale in the absence of the {\it in situ} calibration (provided by the dijet mass templates), by removing the parameterization as a function of JES and by maximizing a likelihood built only with the top templates corresponding to JES = 0 $\sigma_{c}$.
We reconstruct the top quark mass for two samples with $M_{top}$ = 170~GeV/$c^{2}$, but with different values for JES$_{true}$: +1 $\sigma_{c}$, and -1 $\sigma_{c}$, respectively. 
Taking half of the difference between the two reconstructed $M_{t}$ determines the systematic uncertainty due to jet energy scale as 2.2~GeV/$c^{2}$, which is $10\%$ more than in the case of using the {\it in situ} calibration and the JES parameterization.


\section{\label{sec:Res} Results}

Applying the event selection described in Section~\ref{sec:EvSel} to the  multijet data corresponding to an integrated luminosity of 943~pb$^{-1}$, we observe 48 events with one secondary vertex tag and 24 events with at least two secondary vertex tags.
Performing the likelihood maximization and applying the corrections described in Section~\ref{sec:MassFit} for this sample, we measure a top quark mass of 171.1 $\pm$ 3.7~GeV/$c^{2}$ and a value for JES of 0.5 $\pm$ 0.9 $\sigma_{c}$. 

\begin{figure}[!htbp]
\begin{center}
\mbox{(a)}
\includegraphics[width=1.0\linewidth]{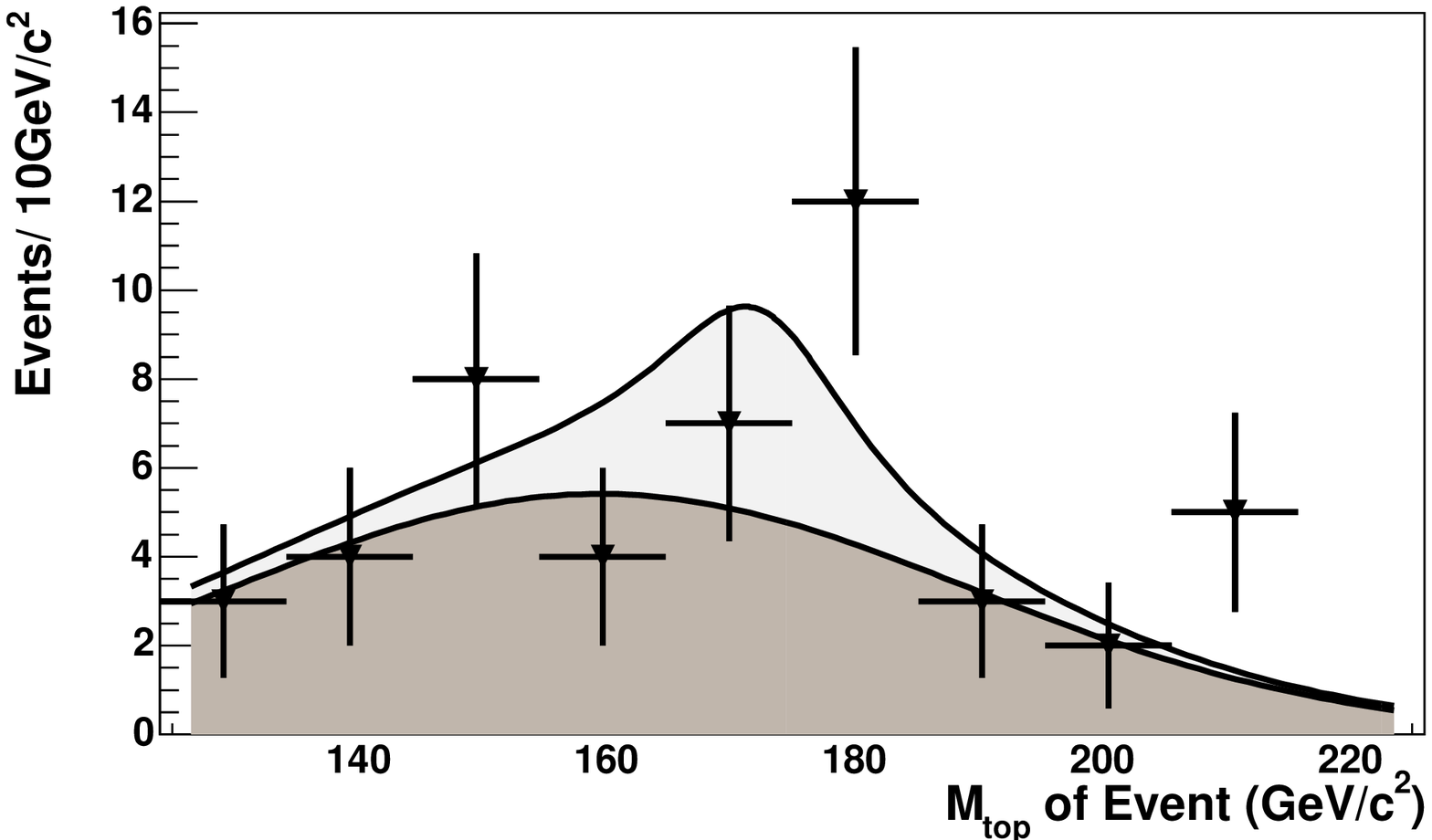}
\mbox{(b)}
\includegraphics[width=1.0\linewidth]{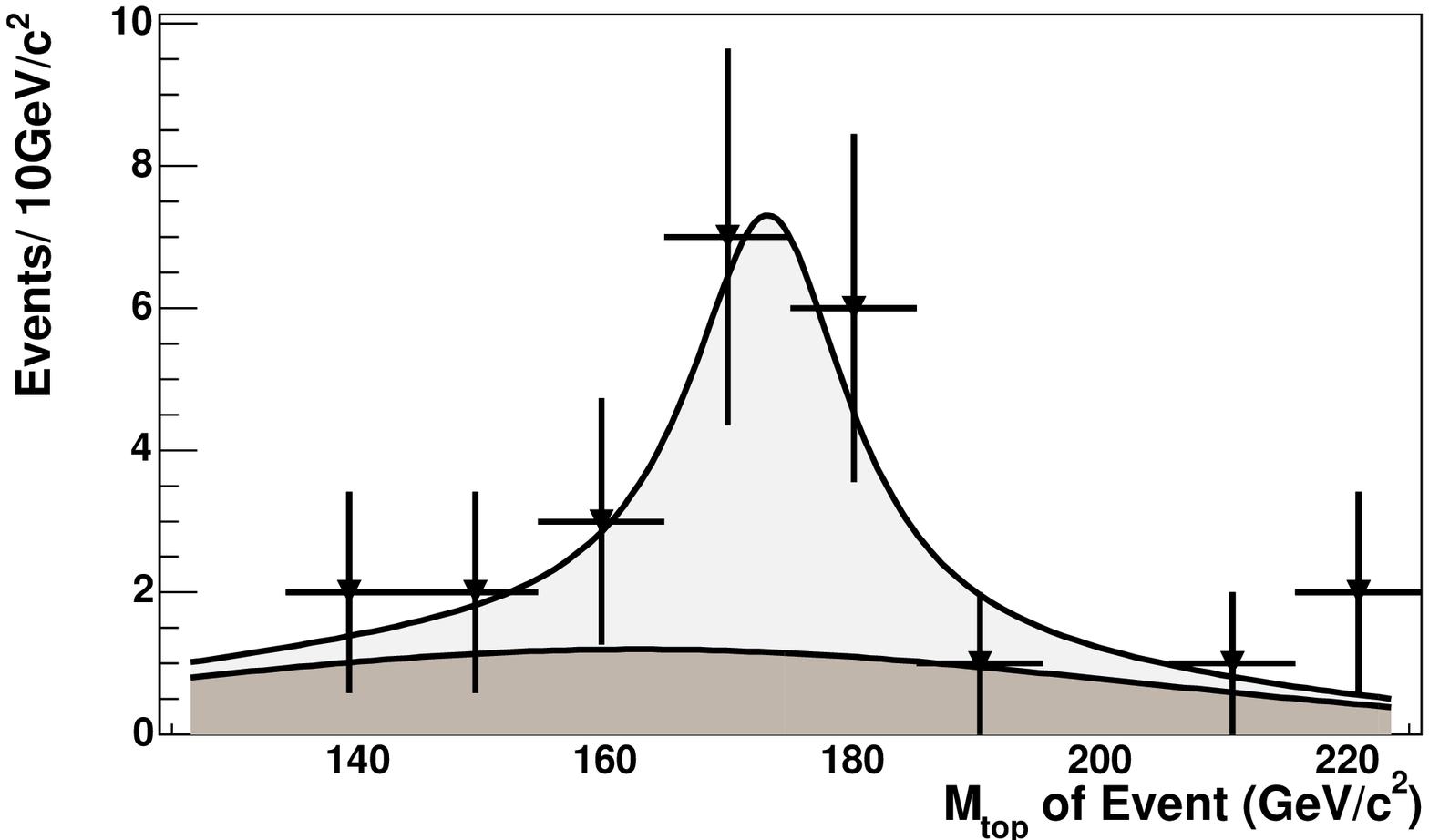}
\caption{Reconstructed top mass for data (points), best signal+background fit (light) and background shape from the best fit (dark): for (a) samplee with only one secondary vertex tag, and (b) the sample with at least two secondary vertex tags.}\label{fig:10-1}
\end{center}
\end{figure}

Figure~\ref{fig:10-1} shows the distributions of reconstructed top quark masses for data (dots) and for the combination (light) of signal and background templates that best fit the data. 
The background (dark) contribution is shown normalized to the data as determined by the fractions obtained from the likelihood fit. 
There are two sets of distributions corresponding to the sample with only one secondary vertex tag (Fig.~\ref{fig:10-1}(a)) and to the sample with at least two secondary vertex tags (Fig.~\ref{fig:10-1}(b)).

\begin{figure}[!htbp]
\begin{center}
\includegraphics[width=\linewidth]{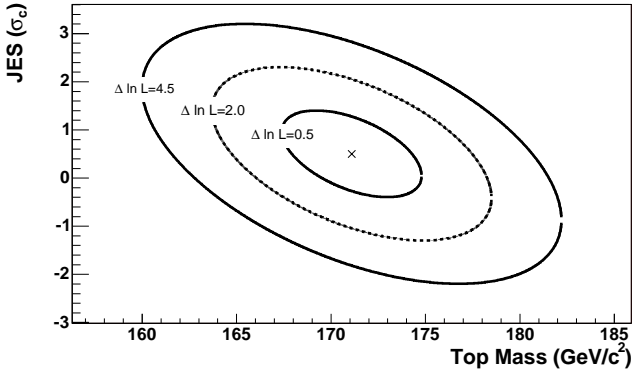}
\end{center}
\caption[Contours of the mass and JES reconstruction in the data]{Contours of the likelihood in the M$_{top}$ and JES plane at a number of values of $\Delta lnL$, the change in negative log-likelihood from its maximum.}\label{fig:10-2}
\end{figure}

The minimized negative log-likelihood is shown in Fig.~\ref{fig:10-2} as a function of the top mass and JES after correcting for bias (Eqs.~\ref{eq:8-6} and~\ref{eq:8-9}) and for non-Gaussian effects (Section~\ref{sec:Check}). 
The central point corresponds to the minimum of the negative log-likelihood, while the contours are given at a number of values of $\Delta$ln$L$, the change in negative log-likelihood from its minimum.

Table~\ref{table:7-1} lists the number of events for \tt and for background for the one- and two-secondary vertex tags cases, as measured following the minimization of the two dimensional likelihood of Eqn.~\ref{eq:7-1} on the data. 

\begin{table}[!htbp]
\begin{center}
\caption{Measured sample composition of the multijet data sample for a luminosity of 943 pb$^{-1}$, passing the event selection. The second column (1 tag) gives the number of events with only one secondary vertex tag, while the third column ($\geq$2 tags) is for the events with at least two secondary vertex tags.}\label{table:7-1}
\begin{tabular*}{\linewidth}{c@{\extracolsep{\fill}}c@{\extracolsep{\fill}}c}
\hline
\hline
Number of Events & 1 tag & $\geq$2 tags\\
\hline
Signal (\tt)  & 13.2 $\pm$ 3.7 & 14.1 $\pm$ 3.4\\
Background  & 34.6 $\pm$ 7.2 & 9.2 $\pm$ 4.3 \\
Total Observed  & 48 & 24 \\
\hline
\hline
\end{tabular*}
\end{center}
\end{table}

\begin{figure}[!htbp]
\begin{center}
\includegraphics[width=\linewidth]{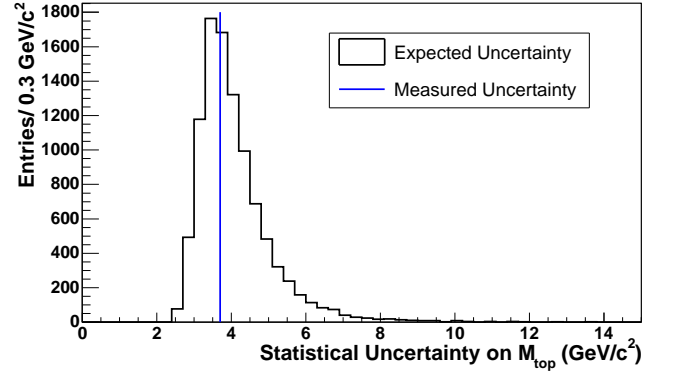}
\end{center}
\caption{Distribution of expected statistical uncertainty on $M_{t}$ (histogram) and the measured uncertainty (vertical line). In about 41$\%$ of simulated experiments a statistical uncertainty on the top quark mass smaller than in the experiment is found.}\label{fig:10-3}
\end{figure}

Using a \tt Monte Carlo sample with a top quark mass equal to 170~GeV/$c^{2}$ and the number of signal and background events from Table~\ref{table:7-1}, we perform simulated experiments and determine the distribution of expected uncertainty on the top quark mass due to statistical effects and JES. 
About $41\%$ of the simulated experiments have a combined uncertainty on the top quark mass lower than the measured value of 3.7~GeV/$c^{2}$. 
This can be seen in Fig.~\ref{fig:10-3}, where the histogram shows the results of the simulated experiments and the vertical line represents the measured uncertainty. 
In conclusion, the measured combined statistical and JES uncertainties on the top mass agree with the expectation.

In order to obtain the contribution of the uncertainty in jet energy scale to the uncertainty on the top quark mass, the minimization of the 2D likelihood is modified such that the JES parameter is fixed to 0.5 $\sigma_{c}$ (the value of JES from the likelihood minimization). 
Following this procedure the uncertainty on the top mass is 2.8~GeV/$c^{2}$. 
Subtracting in quadrature this value from the uncertainty obtained when the JES was not fixed (3.7 GeV/$c^{2}$), we estimate the systematic uncertainty contributed by JES as 2.4~GeV/$c^{2}$.

\section{\label{sec:Syst} Systematic Uncertainties}

We model \tt events using simulated events, which do not always accurately describe all effects we expect to see in the data.
The major sources of uncertainties appear from our understanding of jet fragmentation, our modeling of the radiation from the initial or final partons, and our understanding of the proton and antiproton internal structure. 
Apart from these uncertainties, which are present in most top quark measurements, we also address other issues specific to the present method such as the shape of the background top templates following the correction for \tt content, and the uncertainty in the two dimensional correction of the reconstructed top mass and JES.

\subsection{Systematic uncertainties related to jet energy scale}

\subsubsection{b-jet energy scale}

We study the effect of the uncertainty on the modeling of $b$ quarks due to the uncertainty in the semi-leptonic branching ratio, the modeling of the heavy flavor fragmentation, and due to the color connection effects.

To determine this we reconstruct the top mass in a simulated \tt sample ($M_{top}$ = 175~GeV/$c^{2}$) where we select $b$-jets by matching the $b$ quarks to a jet.
The matching procedure requires $\sqrt{(\Delta \eta)^{2} + (\Delta \phi)^{2}}$ $<$ 0.4 between the quark and the jet.
We modify the energy of the $b$-jets by 0.6$\%$ corresponding to the uncertainty on the $b$-jet energy due to the effects listed above~\cite{bjes}. 
The resulting systematic uncertainty on the top quark mass due to the uncertainty on the $b$-jet energy scale is 0.4~GeV/$c^{2}$.

\subsubsection{Residual jet energy scale}

From the two dimensional fit for mass and JES, we extract an uncertainty on the top quark mass that includes a statistical component as well as a systematic uncertainty due to the uncertainty on the jet energy scale. 
This systematic uncertainty is a global estimate of the uncertainty due to jet energy scale. 
Additional detailed effects arise from the limited understanding of the individual contributions to JES (see Section~\ref{sec:MassFit}). 

For this we have to study the effect on the top mass reconstruction from each of these sources: angular dependence of the calorimeter response, contributions by multiple interactions in the same event bunch, modeling of hadron jets, modeling of the underlying event, modeling of parton showers and energy leakage. 
A simulated \tt sample ($M_{top}$ = 175~GeV/$c^{2}$) is used where the energies of the jets have been shifted up or down by the uncertainty at each level separately. 
We reconstruct the top quark mass for each case, without applying any constraint on the value of JES. 
Table~\ref{table:9-2} shows the average shift on the top mass at each level, and the sum in quadrature of these effects. 
We conclude from this study that the uncertainty on the top quark mass contributed by these corrections to the jet energy is 0.7~GeV/$c^{2}$.

\begin{table}[!htbp]
\begin{center}
\caption{Residual jet energy scale uncertainty on the top mass. The sum in quadrature of all the effects represents the total residual systematic uncertainty due to jet energy scale.}\label{table:9-2}
\begin{tabular*}{\linewidth}{l@{\extracolsep{\fill}}c}
\hline
\hline
Source of Systematic & $\delta$M$_{t}$(GeV/$c^{2}$)\\
\hline
Response Relative to Central Calorimeter & 0.2 \\
Multiple Interactions & 0.1 \\
Modeling of Hadron Jets & 0.5 \\
Modeling of the Underlying Event & 0.0 \\
Modeling of Parton Showers & 0.5 \\
Energy Leakage & 0.1 \\
\\
Total Residual JES Uncertainty & 0.7 \\
\hline
\hline
\end{tabular*}
\end{center}
\end{table}

\subsection{Systematic uncertainties due to background}

\subsubsection{\label{sec:bgshape} Background modeling}

Based on the background model (Section~\ref{sec:Bckd}), we assume $M_{top}$ = 170~GeV/$c^{2}$ to correct for the presence of \tt events in the background distributions. 
To estimate the uncertainty associated with making this assumption, we modify our background model considering a 10~GeV/$c^{2}$ variation on $M_{top}$ used in the default background correction procedure. 
This variation results in a change in the value of the reconstructed top quark mass by 0.9~GeV/$c^{2}$ which is added as a systematic uncertainty.

\subsubsection{Background statistics}

Another effect we address here is that of the limited statistics ($\approx$ 2600 events, see Section~\ref{sec:Bckd}) of the data sample used to model the background. 
To estimate this effect we vary the parameters describing the background templates within their uncertainties. 
Using the procedure described below, we find that the effect on the reconstructed top quark mass due to variation on the background dijet mass templates is negligible. 
This is not the case of the background top templates. 


For simplicity, we label the parameters of this template as Constant, Mean and Sigma, representing the constant, the mean and the width of the Gaussian function describing the background top template. 
In order to find the uncertainties on these parameters, we vary the content of the top template histograms for background assuming that each bin fluctuates according to a Poisson probability.
This variation is done 10,000 times, and each time we extract and form distributions with the values of the three parameters, Constant, Mean and Sigma after applying the correction due to the residual \tt content in the sample.
We use the spread of these distributions as the uncertainties on the parameters of the top templates for background. 

Table~\ref{table:9-1} shows the values of these uncertainties separately for the sample with only one secondary vertex tag (1tag) and for the sample with at least two secondary vertex tags (2tags).
Varying the parameters of the background top templates within these uncertainties results in a shift in the reconstructed top quark mass of 0.4~GeV/$c^{2}$ and we add this as a new systematic uncertainty.

\begin{table}[!htbp]
\begin{center}
\caption{Parameters of the top templates for background events. These templates have been described in Section~\ref{sec:Templ}. The second column is for the single tagged sample (1tag), while the third column is for the double tagged sample (2 tags).}\label{table:9-1}
\begin{tabular*}{\linewidth}{c@{\extracolsep{\fill}}c@{\extracolsep{\fill}}c}
\hline
\hline
Parameter & 1 tag & 2 tags\\
\hline
Constant ((GeV/$c^{2}$)$^{-1}$) & 0.015$\pm$0.001 & 0.013$\pm$0.001 \\
Mean (GeV/$c^{2}$)& 159$\pm$3 & 163$\pm$3 \\
Sigma ((GeV/$c^{2}$)$^{2}$) & 1790$\pm$272 & 3280$\pm$712 \\
\hline
\hline
\end{tabular*}
\end{center}
\end{table}

\subsection{Initial and final state radiation}

The top quark mass measurement is affected by how we model the initial and final state gluon radiation. 
This radiation affects the jet multiplicity in the event as well as the energy of the jets, which in turn affect the top quark mass reconstruction. 

The amount of radiation from the initial partons is controlled in our simulated \tt samples by the DGLAP evolution equation~\cite{ifsrsys1}~\cite{ifsrsys2}. 
The parameters of these equations are $\Lambda_{QCD}$ and K (the scale of the transverse momentum for showering). 
In the case of the initial state radiation, these parameters are tuned in the simulation to reflect the amount of radiation observed in Drell-Yan events~\cite{bjes}. 
The amount of radiation, proportional to the average transverse momentum of the leptons, is found to depend smoothly on the invariant mass of the leptons, over a range of energies extending up to the range of \tt events. 
Two sets of values for the parameters $\Lambda_{QCD}$ and K are determined to cover the variation of this dependence within one standard deviation ($\sigma_{ISR}$).

We generate two samples of \tt events ($M_{top}$ = 178~GeV/$c^{2}$) where the parameters $\Lambda_{QCD}$ and K correspond to $+\sigma_{ISR}$ (increase the amount of radiation), and $-\sigma_{ISR}$ (decrease the amount of radiation), respectively.
Using the default set of values, the reconstructed top quark mass is 178.6~GeV/$c^{2}$. 
For the sample with $+\sigma_{ISR}$ the reconstructed top quark mass is 178.9~GeV/$c^{2}$, and for the one with $-\sigma_{ISR}$ the reconstructed top quark mass is 178.6~GeV/$c^{2}$. 
Taking the maximum change in top mass, we quote 0.3~GeV/$c^{2}$ as the uncertainty due to initial state radiation modeling.

Using the same variation of the parameters $\Lambda_{QCD}$ and K to describe the variation of the final state radiation, we reconstruct the top quark mass to be 177.7~GeV/$c^{2}$ in a sample with increased radiation and 177.4~GeV/$c^{2}$ when we decrease the amount of radiation. 
Taking into account the value of the reconstructed top quark mass in the default case, the maximum change in the reconstructed top quark mass is 1.2~GeV/$c^{2}$ representing the systematic uncertainty on the modeling of the final state radiation.

\subsection{Proton and antiproton PDFs}

In our default simulation, the internal structures of the proton and antiproton are given by the {\sc cteq5l} set of functions, and for a \tt sample with $M_{top}$ = 178~GeV/$c^{2}$ the reconstructed top quark mass is 178.6~GeV/$c^{2}$. 
For the same $M_{top}$ value, using a different set of functions ({\sc cteq6m}) results in a reconstructed top quark mass of 178.7~GeV/$c^{2}$. 
Within the {\sc cteq6m} set, there are 20 independent parameters whose uncertainties are representative of the uncertainty on the modeling of such structure functions~\cite{pdfsys}. 
Adding in quadrature all the 20 offsets observed in top quark mass reconstruction due to these variations, we get 0.4~GeV/$c^{2}$.

Also, it is known that the value of $\Lambda_{QCD}$ has a direct effect on the shape of the structure functions. 
In order to estimate this effect, we chose yet another set of PDFs given by MRST, and reconstructed the top mass for $\Lambda_{QCD}$ = 228~GeV to get a top mass of 177.4~GeV/$c^{2}$, and for $\Lambda_{QCD}$ = 300~GeV to get a top mass of 177.7~GeV/$c^{2}$. 
Therefore the systematic uncertainty due to the value of $\Lambda_{QCD}$ is 0.3~GeV/$c^{2}$.

Adding the two contributions in quadrature, we quote that the total systematic uncertainty due to the choice of structure functions of proton and antiproton is 0.5~GeV/$c^{2}$.

\subsection{Other systematic uncertainties}

The default Monte Carlo generator used to determine our templates is {\sc Herwig}, which is known to differ from the {\sc Pythia} generator. 
For simulated \tt samples with $M_{top}$ = 178~GeV/$c^{2}$, we reconstruct the top quark mass as 177.6~GeV/$c^{2}$ using {\sc Herwig} as the generator, and 178.6~GeV/$c^{2}$ using {\sc Pythia}. 
We assign a systematic uncertainty due to the choice of the Monte Carlo generator of 1.0~GeV/$c^{2}$ representing the difference between the reconstructed top quark masses in {\sc Herwig} and {\sc Pythia}.

In addition, we have varied the parameters of Eq.~\ref{eq:8-6} within their uncertainties as listed in Table~\ref{table:8-2}, and obtained new values of the top quark mass. 
The changes from the default value are within 0.2~GeV/$c^{2}$.

\subsection{Summary of the systematic uncertainties}

The total systematic uncertainty on the top mass combining all the effects listed above is 2.1~GeV/$c^{2}$. Table~\ref{table:9-3} summarizes all sources of systematic uncertainties with their individual contribution as well as the combined effect.

\begin{table}[!htbp]
\begin{center}
\caption{Summary of the systematic sources of uncertainty on the top mass. The sum in quadrature of all the effects represents the total systematic uncertainty.}\label{table:9-3}
\begin{tabular*}{\linewidth}{c@{\extracolsep{\fill}}c}
\hline
\hline
Source & Uncertainty (GeV/$c^{2}$)\\
\hline
$b$-jet JES & 0.4 \\
Residual JES & 0.7 \\
Background Modeling & 0.9 \\
Background Statistics & 0.4 \\
Initial State Radiation & 0.3 \\
Final State Radiation & 1.2 \\
\pp PDF Choice & 0.5 \\
{\sc Pythia} vs. {\sc Herwig} & 1.0 \\
Method Calibration & 0.2 \\

Sample Composition & 0.1 \\

\\
Total & 2.1 \\
\hline
\hline
\end{tabular*}
\end{center}
\end{table}

\section{\label{sec:Concl} Conclusion}

We measure the mass of the top quark to be 171.1~GeV/$c^{2}$ with a total uncertainty of 4.3~GeV/$c^{2}$. 
This measurement, the most precise to-date in the all-hadronic channel, is performed using 943~pb$^{-1}$ of integrated luminosity collected with the CDF II detector. 
This is the first simultaneous measurement of the top quark mass and of the jet energy scale in the \tt all-hadronic channel. 
It is also the first mass measurement in this channel that involved the use of the \tt matrix element in the event selection as well as in the mass measurement itself. 

The previous best mass measurement published in this channel, for an integrated luminosity of 1~fb$^{-1}$, has an equivalent total uncertainty of 5.3~GeV/$c^{2}$~\cite{ahmass2} which is 23$\%$ more than in this measurement. 
The main source for the observed improvement is the reduction of the uncertainty on the top quark mass due to jet energy scale (JES). 
In the present analysis, this uncertainty is 2.5~GeV/$c^{2}$ (including the residual JES uncertainty of 0.7 GeV/$c^{2}$), which is about twice smaller than the corresponding uncertainty of 4.5~GeV/$c^{2}$ determined in Ref.~\cite{ahmass2}.


The top quark mass measured in this analysis is consistent with the most precise top quark mass values measured at the Tevatron and at CDF in the lepton+jets~\cite{lepjets} and the dilepton~\cite{dilepton} channels. 
This consistency among the decay channels restricts the possibility for new physics to prefer the \tt all-hadronic decay channel over the other decay channels. 
Table~\ref{table:10-1} summarizes the most precise top quark mass measurements made at the Tevatron using an integrated luminosity of about 1~fb$^{-1}$. 
From this table it can be seen that the all-hadronic channel provides the second most precise top quark mass measurement.

\begin{table}[!htbp]
\begin{center}
\caption{Most precise results from each \tt decay channel from the Tevatron by March 2007. The integrated luminosity used in these analyses is about 1~fb$^{-1}$.}\label{table:10-1}
\begin{tabular*}{\linewidth}{c@{\extracolsep{\fill}}c}
\hline
\hline
Channel & Result\\
\hline
Lepton+Jets~\cite{lepjets} & 170.9$\pm$2.5~GeV/$c^{2}$\\
Dilepton~\cite{dilepton} & 164.5$\pm$6.5~GeV/$c^{2}$\\
All-hadronic (this analysis) & 171.1$\pm$4.3~GeV/$c^{2}$\\
\hline
All-hadronic (previous result)~\cite{ahmass2} & 174.0$\pm$5.3~GeV/$c^{2}$\\
\hline
\hline
\end{tabular*}
\end{center}
\end{table}


As the luminosity collected with the CDF II detector increases to an expected 7~fb$^{-1}$ for Run II, the statistical uncertainty on the top quark mass will improve and additional top quark mass results from CDF are expected in the near future.
A more careful estimation of the sources of systematic uncertainties on the top quark mass as well as a more efficient \tt event selection can help to further reduce the total uncertainty in this analysis. 
We expect that future mass measurements performed in this channel using an increased data sample size will improve the total uncertainty on the top quark mass which will contribute to our understanding of the electroweak interaction as well as to the search for new physics.

\begin{acknowledgments}
We thank the Fermilab staff and the technical staffs of the participating institutions for their vital contributions. This work was supported by the U.S. Department of Energy and National Science Foundation; the Italian Istituto Nazionale di Fisica Nucleare; the Ministry of Education, Culture, Sports, Science and Technology of Japan; the Natural Sciences and Engineering Research Council of Canada; the National Science Council of the Republic of China; the Swiss National Science Foundation; the A.P. Sloan Foundation; the Bundesministerium f\"ur Bildung und Forschung, Germany; the Korean Science and Engineering Foundation and the Korean Research Foundation; the Science and Technology Facilities Council and the Royal Society, UK; the Institut National de Physique Nucleaire et Physique des Particules/CNRS; the Russian Foundation for Basic Research; the Ministerio de Ciencia e Innovaci\'{o}n, and Programa Consolider-Ingenio 2010, Spain; the Slovak R\&D Agency; and the Academy of Finland. 
\end{acknowledgments}


\begin{thebibliography}{999}
\bibitem{met} The missing transverse energy $\vec{\not\!\!{E_{T}}}$ is calculated as the vector sum of the energy in each calorimeter tower multiplied by a unit vector in the azimuthal direction of the vector. $\not\!\!{E_{T}}$ is defined as the magnitude of $\vec{\not\!\!{E_{T}}}$.
\end{thebibliography}
\end{document}